\documentclass[
reprint,
superscriptaddress,
amsmath,amssymb,
aps,
pra,
floatfix,
]{revtex4-2}

\def\final{0}

\usepackage{braket}
\usepackage{mathtools}
\usepackage{graphicx}
\usepackage{color, xcolor}
\usepackage{dcolumn}
\newcolumntype{z}[1]{D{.}{.}{#1}}
\usepackage{physics}
\usepackage{subfigure}

\usepackage[linesnumbered,ruled ]{algorithm2e}
\SetKwInput{KwOutput}{Output}
\SetKwInOut{Input}{Input}
\SetKwInput{Output}{Output} 
\UseRawInputEncoding 

\usepackage{xargs}
\usepackage[colorinlistoftodos,prependcaption]{todonotes}

\definecolor{Asparagus}{rgb}{0.53, 0.66, 0.42}
\definecolor{cornflowerblue}{rgb}{0.39, 0.58, 0.93}
\definecolor{darkolivegreen}{rgb}{0.33, 0.42, 0.18}
\definecolor{awesome}{rgb}{1.0, 0.13, 0.32}

\newtheorem{definitionenv}{Definition}
\newtheorem{lemmaenv}[definitionenv]{Lemma}
\newtheorem{theoremenv}[definitionenv]{Theorem}
\newtheorem{corollaryenv}[definitionenv]{Corollary}
\newtheorem{propositionenv}[definitionenv]{Proposition}
\newtheorem{conjectureenv}[definitionenv]{Conjecture}
\newtheorem{remarkenv}[definitionenv]{Remark}
\newenvironment{remark}{\begin{remarkenv}\rm}{\end{remarkenv}}
\newcommand{\br}{\begin{remark}}
	\newcommand{\er}{\end{remark}}

\newtheorem{exampleenv}{Example}
\newtheorem{app-lemmaenv}[section]{Lemma}

\newenvironment{definition}{\begin{definitionenv}\rm}{\end{definitionenv}}
\newenvironment{lemma}{\begin{lemmaenv}\rm}{\end{lemmaenv}}
\newenvironment{theorem}{\begin{theoremenv}\rm}{\end{theoremenv}}
\newenvironment{corollary}{\begin{corollaryenv}\rm}{\end{corollaryenv}}
\newenvironment{example}{\begin{exampleenv}\rm}{\end{exampleenv}}
\newenvironment{proposition}{\begin{propositionenv}\rm}{\end{propositionenv}}
\newenvironment{conjecture}{\begin{conjectureenv}\rm}{\end{conjectureenv}}
\newenvironment{app-lemma}{\begin{app-lemmaenv}\rm}{\end{app-lemmaenv}}

\newcommand{\bd}{\begin{definition}}
	\newcommand{\ed}{\end{definition}}
\newcommand{\bl}{\begin{lemma}}
	\newcommand{\el}{\end{lemma}}
\newcommand{\elp}{\hspace*{\fill} $\Box$
\end{lemma}}
\newcommand{\bt}{\begin{theorem}}
\newcommand{\et}{\end{theorem}}
\newcommand{\etp}{\hspace*{\fill} $\Box$
\end{theorem}}
\newcommand{\bc}{\begin{corollary}}
\newcommand{\ec}{\end{corollary}}
\newcommand{\ecp}{\hspace*{\fill} $\Box$
\end{corollary}}
\newcommand{\bcj}{\begin{conjecture}}
\newcommand{\ecj}{\end{conjecture}}

\newcommand{\be}{\begin{example}}
\newcommand{\ee}{\end{example}}
\newcommand{\eep}{\hspace*{\fill} $\Box$
\end{example}}
\newcommand{\bp}{\begin{proposition}}
\newcommand{\ep}{\end{proposition}}
\newcommand{\epp}{
\end{proposition}}

\newcommand{\wt}{\mathrm{wt}}

\ifnum\final=0
\newcommand{\mynote}[2]{{\color{#1} \marginpar{\tiny #2}}}
\newcommand{\mybignote}[2]{{\color{#1} $\langle \langle$ #2$\rangle \rangle$}}
\newcommandx{\rednote}[2][1=]{\todo[linecolor=red,backgroundcolor=red!25,bordercolor=red,#1]{#2}}
\newcommandx{\bluenote}[2][1=]{\todo[linecolor=blue,backgroundcolor=blue!25,bordercolor=blue,#1]{#2}}
\newcommandx{\yellownote}[2][1=]{\todo[linecolor=yellow,backgroundcolor=yellow!25,bordercolor=yellow,#1]{#2}}
\newcommandx{\greennote}[2][1=]{\todo[inline,linecolor=olive,backgroundcolor=green!25,bordercolor=olive,#1]{#2}}

\else
\newcommand{\mynote}[2]{}
\newcommand{\mybignote}[2]{}
\newcommand{\rednote}[2][1=]{}
\newcommand{\bluenote}[2][1=]{}
\newcommand{\greennote}[2][1=]{}
\newcommand{\yellownote}[2][1=]{}

\fi

\usepackage{adjustbox}
\usetikzlibrary{quantikz2}
\usetikzlibrary{backgrounds,fit,decorations.pathreplacing}  
\usetikzlibrary{automata} 
\usetikzlibrary{circuits.logic.CDH}
\usetikzlibrary{circuits.logic.US}
\usetikzlibrary{mindmap}
\usetikzlibrary{decorations}
\usetikzlibrary{decorations.pathmorphing}
\usetikzlibrary{arrows,decorations.pathreplacing}

\tikzset{meter/.append style={draw, inner sep=10, rectangle, font=\vphantom{A}, minimum width=30, line width=.4, path picture={\draw[black] ([shift={(.1,.3)}]path picture bounding box.south west) to[bend left=50] ([shift={(-.1,.3)}]path picture bounding box.south east);\draw[black,-latex] ([shift={(0,.1)}]path picture bounding box.south) -- ([shift={(.3,-.1)}]path picture bounding box.north);}}}

\begin{document}

\title{
High-Rate Amplitude-Damping Shor Codes with Immunity to Collective Coherent Errors
} 

\author{En-Jui Chang}
 \email{ejchang@nycu.edu.tw}
 \affiliation{Institute of Communications Engineering, National Yang Ming Chiao Tung University,  Hsinchu 30010, Taiwan} 
\author{Ching-Yi Lai}
 \email{cylai@nycu.edu.tw}
 \affiliation{Institute of Communications Engineering, National Yang Ming Chiao Tung University,  Hsinchu 30010, Taiwan}
 
\date{\today}
 
\begin{abstract}

We introduce a family of high-rate amplitude-damping (AD) Shor Codes, designed to effectively correct AD errors while maintaining immunity to collective coherent (CC) errors. The proposed $[[(w+1)(w+K), K]]$ AD codes can approximately correct up to $w$ AD errors, with flexible parameters $(w, K)$, and we provide a rigorous proof that these codes satisfy the approximate quantum error correction conditions. These codes leverage structured stabilizer measurements, enabling efficient detection of AD errors using only local operations and ancillary qubits. We further construct a family of CC-AD Shor codes by concatenating these AD codes with the dual-rail code.
\end{abstract}

\maketitle

The Hamiltonian $\hat{H}$ of a quantum system governs its unitary evolution, determining the time dynamics of the quantum state according to Schr\"{o}dinger's equation. Its inherent properties can lead to a specific type of noise, known as coherent errors~\cite{Chamberland2017, Bravyi2018, Debroy2018, Mrton2023}, even in the absence of external interference. For instance, a mismatch between the desired Hamiltonian and the actual one can induce coherent errors on the qubits. These errors are systematic and time-correlated, accumulating over time as the system evolves. A prominent example is collective coherent (CC) errors, where all qubits undergo identical evolution under the same Hamiltonian.

In any realistic quantum system, qubits are not perfectly isolated from their environment and energy spontaneously dissipates into the environment, leading to amplitude-damping (AD) errors~\cite{Gutirrez2013, Piedrafita2017, Grassl2018, Joshi2021}. The rate of amplitude damping $\gamma$ is governed by the relaxation time  $T_{1}$, with $\gamma=1-e^{-(t/T_{1})}$ for a process lasting time $t$. This dissipation affects the fidelity of quantum states during communication and computation. For instance, in long-distance quantum communication networks like quantum key distribution~\cite{BENNETT20147}, AD errors commonly referred to as photon-loss errors become significant as communication distances or times increase.

However, in various contexts such as quantum clock synchronization~\cite{PhysRevLett.85.2010} or quantum repeaters~\cite{Goodenough2016, Rozpdek2021}, AD errors and CC errors are often treated separately. The interplay between coherent evolution and dissipative processes makes it crucial to design robust quantum error correction (QEC) codes~\cite{CS96, Ste96, Knill1997} capable of handling AD and CC errors.

In any quantum communication process of non-negligible duration, accumulated CC errors cannot be effectively mitigated by modeling them as stochastic Pauli noise. This is because the coherent evolution, in which all qubits undergo identical rotations, can induce correlated errors that exceed the code distance of typical stabilizer codes. Similarly, accumulated AD errors deviate fundamentally from stochastic Pauli models: under amplitude damping, an initially maximally mixed state will lose energy and become biased, whereas the Pauli noise model assumes the state remains maximally mixed. These discrepancies highlight the necessity of analyzing CC and AD errors directly, rather than relying on stochastic Pauli approximations commonly assumed in standard QEC frameworks.

QEC codes designed to correct both AD and CC errors have made significant progress in recent years. In particular, constant-excitation (CE) codes, constructed through concatenation of stabilizer codes with the dual-rail code~\cite{Knill2001}, have demonstrated inherent immunity to CC errors~\cite{Ouyang2021, HLRC22}. Ouyang~\cite{Ouyang2021} proposed the $[[8,1,2]]$ CE code, while Hu \textit{et al.}~\cite{HLRC22} introduced the $[[(2L)^2,1,2L]]$ CE codes. The former approach, however, traces back to the work of Duan \textit{et al.}~\cite{Duan2010}, who explored the construction of AD error-correcting codes using similar techniques. Specifically, given an $[[N,K,d]]$ code that encodes $K$ logical qubits into $N$ physical qubits and can detect up to $(d-1)$ Pauli errors, one can construct a $[[2N,K]]$ (CC-)AD code capable of correcting up to $(d-1)$ AD errors. An $[[N,K]]$ quantum code has the code rate $\frac{K}{N}$ representing the number of logical qubits encoded per physical qubit. Such and AD code is referred to as a \textit{$(d-1)$-code}. On the other hand, the latter approach by Hu \textit{et al.} is more appropriately viewed as a modified CE variant of the Shor codes designed to correct AD errors~\cite{Piedrafita2017}.

Although every stabilizer code can be transformed into a CE QEC code~\cite{Duan2010} at the cost of halving its code rate, it remains important to design code families providing higher code rates. In particular, there are two main directions to improve code rates in CE QEC stabilizer codes. The first is to develop outer stabilizer codes with higher code rates, which also improves performance in regimes where only AD errors are considered. The second is to go beyond the construction framework of Duan \textit{et al.}~\cite{Duan2010}. In this work, we focus on the former direction; the improvement of the latter is addressed in our recent work, quantum dual extended Hamming codes $[[2^{r+1}, 2^r-(r+1), 4]]$~\cite{2503.05249}.

To address such limitations in correcting AD errors, we propose an alternative family of $[[(w+1)(w+K), K]]$ AD $w$-codes, generalizing the Shor codes~\cite{Shor95,Got97,Duan2010}. Inspired by the high-rate codes introduced by Fletcher \textit{et al.}~\cite{Fletcher2008}, this code family includes both the Shor codes~\cite{Shor95} and the codes defined by Fletcher \textit{et al.}~\cite{Fletcher2008} as special cases.

Our code family inherits the benefits of the Shor codes and is constructed as a Calderbank-Shor-Steane (CSS) stabilizer code~\cite{CS96,Ste96}, enabling transversal logical Pauli gates, i.e., any logical Pauli operator $\bar{P}$ corresponds to a tensor product of physical Pauli operators $\hat{P}$, where $P \in {X, Y, Z}$. These transversal gates keeping errors from accumulating are highly desired in fault-tolerant QEC~\cite{PhysRevLett.111.090505}. An additional motivation for extending the Shor code family is that several of these codes have corresponding bosonic versions, as discussed in our recent work, high-rate extended binomial codes~\cite{2501.07093}.

We also provide a rigorous proof of approximate QEC condition for these codes. These codes feature explicit, simple encoding circuits, efficient logical operations, and high code rates. By concatenating these AD codes with the dual-rail code, we further construct a corresponding family of CC-AD codes. Note that this paper does not consider the generalized Bacon-Shor subsystem codes~\cite{Bacon05, Bravyi2011}.

{\textit{(Composite error model.)}}
The time evolution of a quantum state in an open system involves the intrinsic Hamiltonian $\hat{H}_{o}$ and energy-dissipation, modeled by coherent and AD errors, respectively. We consider an $N$-qubit quantum system to be affected by both types of errors.

We assume that each qubit evolves under the same intrinsic Hamiltonian. Without loss of generality, let $\hat{H}_{o}\coloneqq g\hat{Z}$, where $g\in\mathbb{R}^{-}$ is a negative coupling and  $\hat{Z}\coloneqq\ket{0}\bra{0}-\ket{1}\bra{1}$ is the Pauli $Z$-operator. The CC error channel is then described by the unitary evolution $\rho\mapsto\hat{U}_{\textrm{CC}}(\Delta t)\rho\hat{U}_{\textrm{CC}}^{\dagger}(\Delta t)$, where
\begin{align}
    \hat{U}_{\textrm{CC}}(\Delta t)\coloneqq \bigotimes_{j=0}^{N-1}e^{-\mathrm{i}\hat{H}_{o,j}\Delta t}
\end{align}
for a positive circuit time $\Delta t\in\mathbb{R}^{+}$ and the subscript of an operator indicates the qubit to which it applies.

We assume that each qubit undergoes an independent AD channel, defined by the Kraus operators $\hat{A}_{0}\coloneqq\ket{0}\bra{0}+\sqrt{1-\gamma}\ket{1}\bra{1}$ and $\hat{A}_{1}\coloneqq\sqrt{\gamma}\ket{0}\bra{1}$, where $\gamma$ is the probability that an excited state $\ket{1}$ relaxes to the ground state $\ket{0}$. Define $\hat{\mathcal{A}}_{a}\coloneqq \hat{A}_{a_{0}}\otimes\hat{A}_{a_{1}}\otimes\cdots\otimes\hat{A}_{a_{N-1}}$ for $a=a_{0}a_{1}\dots a_{N-1}\in\{0,1\}^{N}$. Let $\wt(a)$ denote the Hamming weight of a binary string $a$, defined as the  number of its nonzero elements. Then, the error operator $\hat{\mathcal{A}}_{a}$ is a weight-$w$ AD error if $w=\wt(a)$, meaning that $w$ qubits undergoing $\hat{A}_{1}$ errors, while the remaining qubits experience $\hat{A}_{0}$ errors.

A CE state in a multi-qubit system is a superposition of states with a fixed number of qubits in the excited state $\ket{1}$. In particular, a CE state that undergoes independent AD errors remains a CE state. For instance, consider $\ket{\psi}=\frac{1}{\sqrt{3}}\left(\ket{110}+\ket{101}+\ket{011}\right)$. Then $\hat{A}_{1}\otimes \hat{A}_{0}\otimes \hat{A}_{0} \ket{\psi} = \sqrt{\frac{\gamma(1-\gamma)}{3}} \left(\ket{010}+\ket{001}\right)$,  which is still a CE state.

Furthermore, a CE state is an eigenstate of $\hat{U}(\Delta t)$ and therefore immune to CC errors. Consequently, the order in which a CE state undergoes CC and AD errors is inconsequential. Thus, the overall error model on an $N$-qubit system is given by  
\begin{align}
    \mathcal{E}^{\textrm{CC-AD}}(\rho)=&\sum_{a\in\{0,1\}^{N}}\hat{\mathcal{A}}_{a}\hat{U}_{\textrm{CC}}(\Delta t)\rho\hat{U}_{\textrm{CC}}^{\dagger}(\Delta t)\hat{\mathcal{A}}_{a}^{\dagger}
\end{align}
for some positive $\Delta t$.

Although CE states avoid CC errors, designing the outer stabilizer code to correct AD errors presents more challenges than addressing stochastic Pauli errors. Unlike Pauli noise, AD error introduces asymmetry in eigenstates, rendering $\hat{X}$-based stabilizer measurements unreliable. A proper QEC stabilizer should satisfy two necessary properties:
\begin{enumerate}
    \item Every codeword must be an eigenstate of the stabilizer with eigenvalue \( +1 \);
    \item Every stabilizer must either commute or anti-commute with each element in the error set.
\end{enumerate}

The first property is straightforward and independent of error models. However, the second property reveals a critical limitation: some stabilizer codes constructed to correct stochastic Pauli errors may not stabilize codewords under AD noise. To illustrate this distinction, consider the transversal stabilizers \(\hat{X}\hat{X}\) and \(\hat{Z}\hat{Z}\) of the Bell state $\ket{\Phi^{+}} = \frac{1}{\sqrt{2}}(\ket{00} + \ket{11})$, which can detect a single Pauli error. These stabilizers either commute or anti-commute with Pauli errors as expected. However, under AD errors, only the \(\hat{Z}\hat{Z}\) stabilizer retains this behavior. The \(\hat{X}\hat{X}\) ``stabilizer,'' on the other hand, fails to stabilize the state affected by AD noise. For instance, 
\begin{align*}
    (\hat{X}\hat{X})(\hat{A}_{0}\otimes\hat{A}_{0})\ket{\Phi^{+}}=&(\hat{X}\hat{X})(\frac{1}{\sqrt{2}}(\ket{00} + (1-\gamma)\ket{11})),\\
    =&(\frac{1}{\sqrt{2}}(\ket{11} + (1-\gamma)\ket{00})).
\end{align*}

As a result, we restrict our measurements to transversal $\hat{Z}$ stabilizers when detecting AD errors. For completeness, we also provide the full set of transversal $\hat{X}$ or $\hat{Z}$ stabilizers for users who may wish to apply our QEC codes in settings involving stochastic Pauli noise.

{\textit{(AD QEC codes.)}}
A quantum code is said to correct weight-$w$ AD errors if it improves the fidelity of the raw transmission value of $1 - O(\gamma)$ to $1 - O(\gamma^{w+1})$. This code is referred to as a $w$-code~\cite{Duan2010}.

We first propose a family of $[[(w+1)(w+K), K]]$ AD $w$-codes, derived from Shor's $[[(w+1)^2, 1]]$ $w$-codes~\cite{Shor95,Got97}, but with higher code rates. The enhanced ability of the Shor code family to correct higher-weight AD errors is attributed to the use of larger inner and outer repetition codes in its construction~\cite{Got97}.

\begin{figure}[htbp]
    \centering
    \subfigure{\includegraphics[scale=0.06]{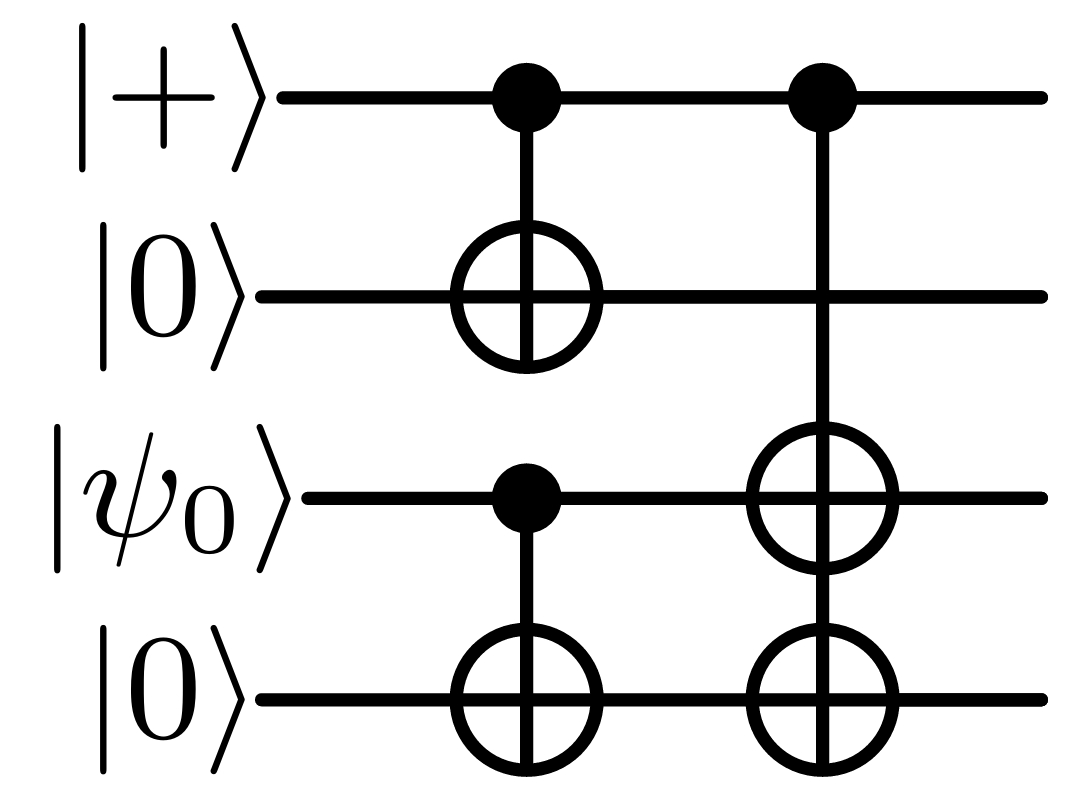}}\label{fig:enc[[4,1]]}
    \subfigure{\includegraphics[scale=0.12]{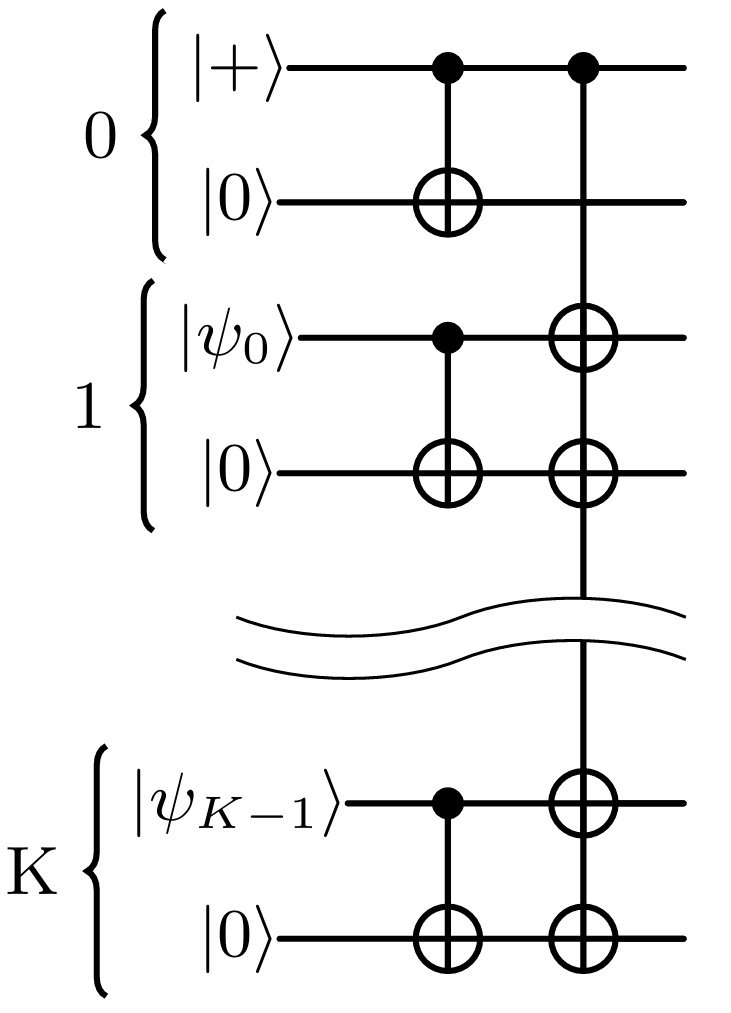}}\label{fig:enc[[2(K+1),K]]}
    \subfigure{\includegraphics[scale=0.12]{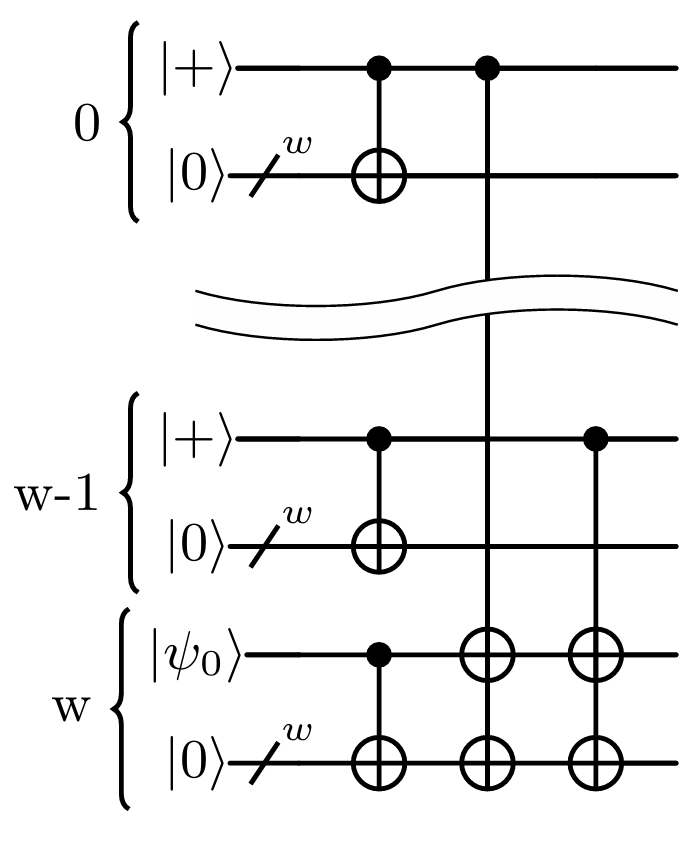}}\label{fig:enc[[(w+1)^{2},1]]}
    \caption{Encoding circuit for (a) the smallest 4-qubit AD QEC 1-code (b) the high rate $[[2(K+1),K]]$ AD $1$-codes with an asymptotic code rate of $\frac{1}{2}$ (c) the $[[(w+1)^{2},1]]$ AD codes capable of protecting against high-weight AD errors. These circuits encode input qubit(s) (a) one qubit $\ket{\psi_{0}}$ into four qubits, (b) $K$ qubits $\ket{\psi_{0}}, \dots, \ket{\psi_{K-1}}$ into $2(K+1)$ qubits, and (c) one qubit $\ket{\psi_{0}}$ into $(w+1)^{2}$ qubits, enabling correction of (a,b) weight-$1$ (c) weight-$w$ AD errors, respectively.
    }
    \label{fig:enc_early}
\end{figure}

Currently, no three-qubit AD $1$-code is known~\cite{OL22}. The $[[4,1]]$ code~\cite{Leung1997} represents the smallest known $1$-code and is also the smallest instance from the $[[(w+1)^2, 1]]$ Shor code family~\cite{Shor95}. Building on this code structure, Fletcher \textit{et al.}~\cite{Fletcher2008} proposed a $[[2(K+1), K]]$ $1$-code family with higher code rates, achieved by asymmetrically concatenating inner and outer codes. This approach involves using multiple inner codes while retaining a single outer code. The encoding circuits for these codes are provided in the Fig.~\ref{fig:enc_early}.

Since QEC schemes for quantum communication must account for various factors, such as the characteristics of the error channel, focusing solely on codeword design does not provide a fair basis for direct comparison. While Fletcher et al.~\cite{Fletcher2008} and Piedrafita and Renes~\cite{Piedrafita2017} both address AD errors, they each approximate these errors differently, making their approaches inherently distinct from ours.

Specifically, Fletcher et al.~\cite{Fletcher2008} model the no-damping event $\hat{A}_{0}$ as equivalent to the identity operator $\hat{I}$, treating it as a no-error event with negligible higher-order distortions. The damping event $\hat{A}_{1}$ is then regarded as the primary perturbation. In contrast, Piedrafita and Renes\cite{Piedrafita2017} approximate AD errors as stochastic Pauli noise, expressed as
\begin{align}
    \mathcal{E}^{\textrm{AD}}_{\textrm{Pauli}}(\rho)=&p_0\rho+p_1\hat{X}\rho\hat{X}+p_2\hat{Y}\rho\hat{Y}+p_3\hat{Z}\rho\hat{Z},
\end{align}
with coefficients $p_0 = 1 - \frac{\gamma}{2} - \frac{\gamma^2}{16}$, $p_1 = p_2 = \frac{\gamma}{4}$, and $p_3 = \frac{\gamma^2}{16}$.
The deviation introduced by this approximation becomes evident when comparing the output states element-wise. For an input state 
\begin{align*}
    \rho=\rho_{0,0}\ket{0}\bra{0}+\rho_{0,1}\ket{0}\bra{1}+\rho_{1,0}\ket{1}\bra{0}+\rho_{1,1}\ket{1}\bra{1},
\end{align*}
the output under the original AD channel is
\begin{align*}
    &\hat{A}_{0}\rho\hat{A}_{0}^{\dagger}+\hat{A}_{1}\rho\hat{A}_{1}^{\dagger}\\
    =&(\rho_{0,0}+\gamma\rho_{1,1})\ket{0}\bra{0}+\sqrt{1-\gamma}\rho_{0,1}\ket{0}\bra{1}\\
    &+\sqrt{1-\gamma}\rho_{1,0}\ket{1}\bra{0}+(1-\gamma)\rho_{1,1}\ket{1}\bra{1},
\end{align*}
whereas the Pauli-approximated output becomes
\begin{align*}
    &\mathcal{E}^{\textrm{AD}}_{\textrm{Pauli}}(\rho)\\
    =&((1-\frac{\gamma}{2})\rho_{0,0}+\frac{\gamma}{2}\rho_{1,1})\ket{0}\bra{0}+(1-\frac{\gamma}{2}-\frac{\gamma^{2}}{8})\rho_{0,1}\ket{0}\bra{1}\\
    &+(1-\frac{\gamma}{2}-\frac{\gamma^{2}}{8})\rho_{1,0}\ket{1}\bra{0}+((1-\frac{\gamma}{2})\rho_{1,1}+\frac{\gamma}{2}\rho_{0,0})\ket{1}\bra{1}.
\end{align*}
For instance, taking the maximally mixed state $\rho_{\textrm{mixed}}=\frac{1}{2}(\ket{0}\bra{0}+\ket{1}\bra{1})$ as the input, the energy-dissipated output state becomes $\rho_{\textrm{out}}=\frac{1+\gamma}{2}\ket{0}\bra{0}+\frac{1-\gamma}{2}\ket{1}\bra{1}$, while the Pauli-approximated output retains the maximally mixed state.

While prior works employ approximations that deviate from the true AD error model, our approach retains the original channel definition. This allows for a more rigorous evaluation of QEC performance against genuine AD errors. Moreover, the choice of error model significantly influences the design of syndrome measurement and recovery operations. As discussed earlier, stabilizers involving $\hat{X}$ operators are ill-suited for true AD error, yet this limitation is often overlooked in QEC schemes that rely on approximated AD models. Consequently, these prior methods require more rigorous validation under realistic noise assumptions.

\begin{figure}[htbp]
    \centering
    \includegraphics[scale=0.25]{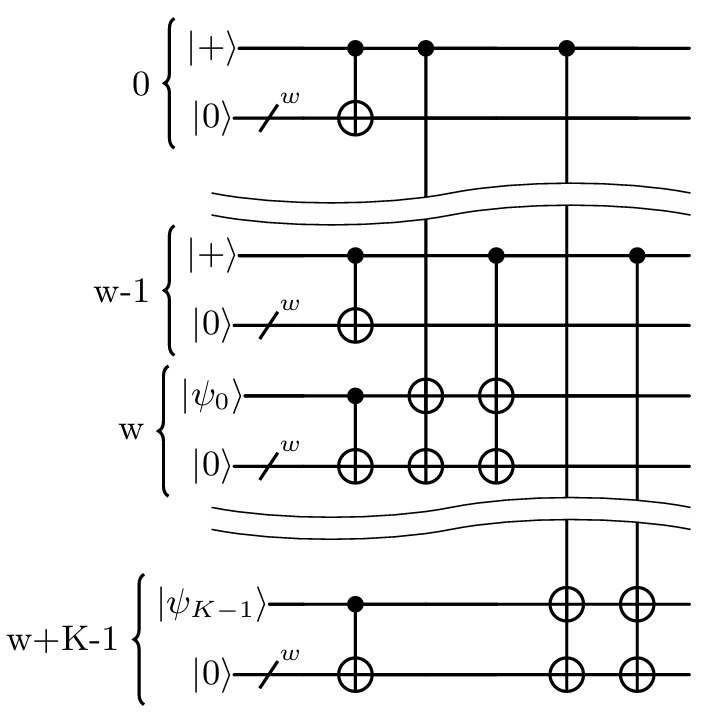}
    \caption{Encoding circuit for the $[[(w+1)(w+K),K]]$ AD Shor codes.
    It encodes $K$ input qubits, $\ket{\psi_{0}}, \dots, \ket{\psi_{K-1}}$, into $(w+1)(w+K)$ qubits to protect against up to weight-$w$ AD errors. 
}\label{fig:enc[(w+1)(w+K),K]}
\end{figure}

By integrating these ideas and restrict the approximation deviated from true AD errors, we construct a unified family of AD Shor codes with the encoding circuit shown in Fig.~\ref{fig:enc[(w+1)(w+K),K]}. 

{\textit{(Approximate quantum error correction conditions.)}}
For the special case $K=1$, our code reduces to the $[[(w+1)^{2},1]]$ $w$-code, which is a symmetric Shor code. 
The Appendix~\ref{APP:AQEC[[(w+1)^2,1]]} provides the conditions for approximate quantum error correction (AQEC) for an $[[(w+1)^2,1]]$ Shor code, with a rigorous proof given. This proof can be extended to our $[[(w+1)(w+K), K]]$ codes as well. We can show that a $[[(w+1)^2,1]]$ code can approximately correct an error set involving AD errors of weight up to $w$ by verifying the Knill-Laflamme QEC conditions~\cite{Knill1997, Got97}.

Similarly, we establish the following AQEC condition for our code family, with the proof deferred to the Appendix~\ref{APP:AQEC[[(w+1)(w+K),K]]}.

\begin{theorem}\label{Thm:AQEC[[(w+1)(w+K),K]]} (AQEC conditions for the $[[(w+1)(w+K),K]]$ $w$-code.)
Let $\big\{\ket{i}_{\textrm{AD}}^{(w,K)}| i \in\{0,1\}^{K}\big\}$ be its codewords shown in the Appendix~\ref{APP:AQEC[[(w+1)(w+K),K]]}. Define the correctable error set $\mathcal{A} = \big\{\hat{\mathcal{A}}_{a} : a \in \{0,1\}^{(w+1)(w+K)}, \wt(a) \leq w \big\}$, where $\wt(k)$ denotes the Hamming weight of $k$. Then, for any $\hat{\mathcal{A}}_{k}, \hat{\mathcal{A}}_{\ell}\in\mathcal{A}$,  we have
\begin{align}
{\bra{i}}_{\textrm{AD}}^{(w,K)}\hat{\mathcal{A}}_{k}^{\dagger}\hat{\mathcal{A}}_{\ell}{\ket{j}}_{\textrm{AD}}^{(w,K)}
=&\delta_{ij}C_{k\ell}+O(\gamma^{w+1}),
\end{align}
where $\delta_{ij}$ is the Kronecker delta function, and $C_{k\ell}$ is a complex number independent of these two length-$K$ binary strings $i$ and $j$.
\end{theorem}

{\textit{(Local stabilizers of the $[[(w+1)(w+K),K]]$ $w$-code.)}}
Our code family inherits the benefits of Shor codes and is composed of Calderbank-Shor-Steane (CSS) stabilizer codes~\cite{CS96,Ste96}. To obtain the stabilizers for the $[[(w+1)(w+K), K]]$ $w$-code, we decompose its codewords into $(w+K)$ blocks, each containing $(w+1)$ physical qubits. Within the $i$-th block for $i \in \{0, \dots, w+K-1\}$, the inner repetition code guarantees the presence of $w$ stabilizers $\hat{S}^{Z}_{i,j} = \hat{Z}_{(w+1)i+j}\hat{Z}_{(w+1)i+j+1}$ for $j \in \{0, \dots, w-1\}$. On the other hand, the outer repetition code ensures the existence of $(w-1)$ stabilizers $\hat{S}^{X}_{i} = \bigotimes_{j=0}^{w}\hat{X}_{(w+1)i+j}\hat{X}_{(w+1)(i+1)+j}$ for $i \in \{0, \dots, w-2\}$, as well as a stabilizer $\hat{S}^{X}_{w-1} = \bigotimes_{i=w-1}^{w-1+K}\bigotimes_{j=0}^{w}\hat{X}_{(w+1)i+j}$, where $\hat{X} \coloneqq \ket{0}\bra{1} + \ket{1}\bra{0}$ is the Pauli $X$-operator.

We emphasize that our QEC strategy involves measuring only $\hat{Z}$-type stabilizers $\hat{S}^{Z}_{i,j}$, a point elaborated upon earlier. This is a departure from standard QEC, where all $N-K$ stabilizers of an $[[N,K,d]]$ code are measured, and also from certain prior works on AD error correction~\cite{Fletcher2008,Piedrafita2017}. This distinction is crucial because AD errors result in a single-bit error syndrome, unlike stochastic Pauli errors which require a two-bit syndrome. Consequently, the inclusion of unnecessary stabilizer measurements in some earlier AD QEC schemes negatively impacted their performance.

{\textit{(Logical operators of the $[[(w+1)(w+K),K]]$ $w$-code.)}}
For the $K$ logical qubits, we denote the logical Pauli operators as $\bar{X}_{\ell}$ and $\bar{Z}_{\ell}$ for $\ell\in\{0,\dots,K-1\}$ and $\bar{Y}_{\ell} \coloneqq -\mathrm{i}\bar{Z}_{\ell}\bar{X}_{\ell}$. Similarly, the logical Hadamard operators are denoted as $\bar{H}_{\ell}\coloneqq\frac{1}{\sqrt{2}}(\bar{X}_{\ell}+\bar{Z}_{\ell})$. These logical $\bar{X}$ and $\bar{Z}$ operators can be transversely implemented by $(w+1)$-local operators as follows:
\begin{align}
    \bar{X}_{\ell} =& \bigotimes_{j=0}^{w}\hat{X}_{(w+1)(w+\ell)+j},\\
    \bar{Z}_{\ell} =& \bigotimes_{i=0}^{w-1}\hat{Z}_{(w+1)i}\otimes\hat{Z}_{(w+1)(w+\ell)}.
\end{align}

In addition, the global logical $\bar{X}_{\textrm{all}} \coloneqq \bigotimes_{\ell=0}^{K-1} \bar{X}_{\ell}$ can be transversely implemented efficiently due to the $X$-type stabilizers.
\begin{align}
    \bar{X}_{\textrm{all}} =& \bigotimes_{j=0}^{w}\hat{X}_{j}.
\end{align}
This efficient global operator $\bar{X}_{\textrm{all}}$ requires only $(w+1)$ local $\hat{X}$ operators, instead of the $(w+1)K$ $\hat{X}$ operators acting. When $K$ is much larger than $(w+1)$, this global operator $\bar{X}_{\textrm{all}}$ is even more efficient than individually flipping these $K$ qubits, provided the cost of encoding and decoding is negligible.

\begin{figure}
    \centering
    \includegraphics[width=0.4\linewidth]{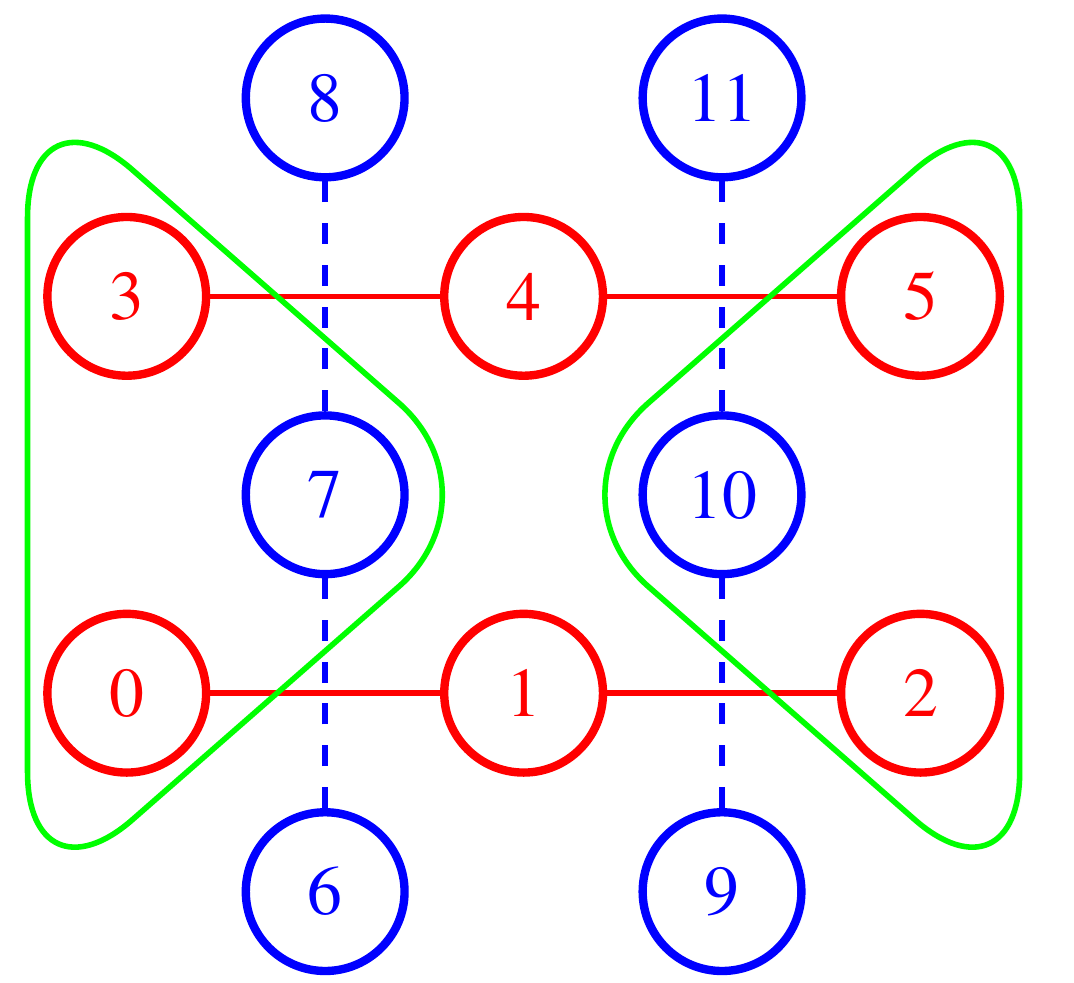}
    \caption{Layout of the $[[12,2]]$ $2$-code.
    In this layout, each solid or dashed edge represents a $Z$ stabilizer, which can be measured locally. The blue chains denote logical $X$ operators, while the red chains correspond to the global logical $X$ operator. Additionally, each red chain, when combined with all the blue chains, forms an $X$ stabilizer. Finally, the two green circles correspond to two distinct logical $Z$ operators.
    }
    \label{fig:1222code}
\end{figure}
 
{\textit{(Two-dimensional layout.)}}
Our AD Shor codes feature a convenient 2D layout. Figure~\ref{fig:1222code} illustrates a possible 2D layout for the $[[12,2]]$ 2-code, with qubits numbered from 0 to 11. It has stabilizers
\begin{align*}
    &Z_0Z_1, \quad Z_1Z_2, \quad Z_3Z_4, \quad Z_4Z_5,\\
    &Z_6Z_7, \quad Z_7Z_8, \quad Z_9Z_{10}, \quad Z_{10}Z_{11},\\
    &X_0X_1X_2X_6X_7X_8X_9X_{10}X_{11},\\
    &X_3X_4X_5X_6X_7X_8X_9X_{10}X_{11}
\end{align*}
and logical operators
\begin{align*}
    &Z_0Z_3Z_7, \quad Z_2Z_5Z_{10}, \quad X_6X_7X_8, \quad X_9X_{10}X_{11}.
\end{align*}
As discussed earlier, the two $\hat{X}$-type stabilizers are not necessary to be measured. However, one possibility for these two remaining stabilizers to become useful is to measure them after the syndrome extraction and recovery of AD errors for a double check, and then correct any noise that occurred during the previous measurements and recovery.

To construct AD Shor codes with more logical qubits, additional blue chains of qubits can be incorporated into the layout. Similarly, to enhance the AD error correction capability, the red chains of qubits can be extended, and additional red chains can be incorporated.

{\textit{(Syndrome extraction and recovery operations.)}}
The stabilizer measurements of the $[[(w+1)(w+K), K]]$ $w$-code can be implemented using CNOTs and ancillary qubits. An AD error event can be detected by measuring $Z$ stabilizers, since $\hat{A}_{0}=\frac{1}{2}(\hat{I}+\hat{Z})+\frac{\sqrt{1-\gamma}}{2}(\hat{I}-\hat{Z})$, $\hat{A}_{1}=\frac{\sqrt{\gamma}}{2}(\hat{X}+\mathrm{i}\hat{Y})$, and  the relations
\begin{align}
    \hat{Z}\hat{A}_{0} = \hat{A}_{0}\hat{Z},\qquad
    \hat{Z}\hat{A}_{1} =& -\hat{A}_{1}\hat{Z}.
\end{align}
In each of the $(w+K)$ blocks, there are $w$ $\hat{Z}^{\otimes 2}$ stabilizers, allowing us to construct a lookup table that maps a length-$w$ binary syndrome string to identify AD errors.

Following the recovery operation detailed in~\cite{Leung1997}, the state can be effectively recovered using the artificial AD channel that counteracts the $\hat{Y}$ error introduced by the natural AD process. A general $\hat{Y}$ rotation with angle $\theta$ is expressed as:
\begin{align}
e^{-\mathrm{i}\frac{\theta}{2}\hat{Y}} &= \cos{\frac{\theta}{2}}\hat{I} - \mathrm{i}\sin{\frac{\theta}{2}}\hat{Y}.
\end{align}

To generate an artificial AD channel, the controlled $\hat{Y}$ rotation with the rotation angles determined by the AD rate $\gamma$ is required. The controlled $\hat{Y}$ rotation with angle $\theta$ can then be written as
\begin{align}
\hat{\textrm{CY}}(\theta)=\ket{0}\bra{0}\otimes\hat{I} + \ket{1}\bra{1}\otimes e^{-\mathrm{i}\frac{\theta}{2}\hat{Y}}.
\end{align}
This controlled $\hat{Y}$ rotation is applied to an input state $\ket{\psi}$ with an ancillary state $\ket{0}$, followed by a measurement of the ancilla in the $Z$-basis. The outcomes correspond to artificial no-damping ($\hat{A}'_{0}$) when the ancilla is measured in $\ket{0}$, and a damped event ($\hat{A'}_{1}$), followed by a conditional application of the Pauli $\hat{X}$ correction on the resulting state. Mathematically:
\begin{align}
(\hat{I}\otimes\bra{0})\hat{\textrm{CY}}(\theta)(\ket{\psi}\otimes\ket{0}) &= (\ket{0}\bra{0}+\cos{\frac{\theta}{2}}\ket{1}\bra{1})\ket{\psi}, \\
\hat{X}(\hat{I}\otimes\bra{1})\hat{\textrm{CY}}(\theta)(\ket{\psi}\otimes\ket{0}) &= (\sin{\frac{\theta}{2}}\hat{Y}\ket{0}\bra{1})\ket{\psi}.
\end{align}
By setting $\sqrt{1-\gamma'} = \cos{\frac{\theta}{2}}$ and $\sqrt{\gamma'} = \sin{\frac{\theta}{2}}$, this procedure effectively implements an artificial AD channel with a damping rate $\gamma' = (\sin{\frac{\theta}{2}})^{2}$. This inherent connection between an undesired $\hat{A}_{0}$ event and a controlled $\hat{Y}$ rotation motivates the use of such rotations for recovering the unwanted phase. A general formulation and a detailed example of this decoding procedure are provided in Appendices~\ref{APP:decode[[(w+1)(w+K),K]]} and~\ref{APP:6_2_1_code}, respectively.

The code rate of this code family asymptotically goes to $\frac{1}{w+1}$ as $K$ approaches infinity, representing a significant square-root improvement over the previous $[[(w+1)^2, 1]]$ $1$-code. Table~\ref{table:AD_QECCs} summarizes the parameters of the AD codes discussed.

\begin{table}[ht]
\begin{tabular}{||c | c | c ||}
 \hline
 \textbf{AD QEC codes} & \textbf{correction weight} & \textbf{code rate}\\ 
 \hline
 \hline
 $[[4,1]]$~\cite{Leung1997} &1 &$\frac{1}{4}$\\
 \hline
 $[[2(K+1),K]]$~\cite{Fletcher2008} &1 &$ \frac{1}{2}\frac{K}{K+1}$\\
 \hline
 $[[(w+1)^{2},1]]$~\cite{Piedrafita2017} &$w$ &$\frac{1}{(w+1)^{2}}$\\
 \hline
 $[[(w+1)(w+K),K]]$ &$w$ &$\frac{1}{w+1}\frac{K}{K+w}$\\
 \textbf{(This work)} & &\\
 \hline
\end{tabular}
\caption{Comparison of different AD QEC codes. The correction weight $w$ of an AD QEC code indicates the maximum weight $w$ of AD errors that can be corrected. }
\label{table:AD_QECCs}
\end{table}

\begin{table}[ht]
\begin{adjustbox}{width=\columnwidth,center}
\begin{tabular}{||c | c | c | c | c | c | c ||} 
 \hline
 \textrm{w\textbackslash K} & $1$ & $2$  & $3$ & $4$ & $5$ & $6$\\ [0.5ex] 
 \hline
 \hline
 $1$ & $(4^{}, 8)$ & $(6^{}, 8)$ & $(8^{}, 12)$ & $(10^{}, 12)$ & $(12^{}, 16)$ & $(14^{}, 16)$\\  
 \hline
 $2$ & $(9^{}, 10)$ & $(12^{}, 16)$ & $(15^{}, 16)$ & $(18^{}, 20)$ & $(21^{}, 22)$ & $(24, 24)$\\  
 \hline
 $3$ & $(16^{}, 20)$ & $(20, 20)$ & $(24, 24)$ & $(28, 24)$ & $(32, 28)$ & $(36, 28)$\\  
 \hline
\end{tabular}
\end{adjustbox}
\caption{The required number of physical qubits $(N_1, N_2)$ for the $[[N_1, K]]$ $w$-codes in this work and the $[[N_2, K]]$ $w$-codes in~\cite{Duan2010} are compared. 
If $N_1 \leq N_2$, it means that our codes require fewer qubits than those in~\cite{Duan2010}.
}
\label{table:Comparison}
\end{table}

In addition, the code rate of this code family outperforms the codes proposed by Duan \textit{et al.}~\cite{Duan2010} for several cases of $(w, K)$, where optimal outer $[[N, K, d]]$ stabilizer codes~\cite{Grassl:codetables, Grassl06, Magma, Brouwer98} are used in the construction by Duan \textit{et al.}.
Table~\ref{table:Comparison} compares the number of qubits required in this work with that in~\cite{Duan2010}. In 12 out of 18 cases shown in Table~\ref{table:Comparison}, our code family requires fewer physical qubits to encode $K$ logical qubits against $w$ AD events.

{\textit{(CE AD QEC codes.)}}
Next, we construct the CE version of the AD QEC code by concatenating AD QEC codes with an inner dual-rail code.
The smallest example is the $[[8,1]]$ $1$-code that comes from concatenating the outer $[[4,1]]$ $1$-code with an inner dual-rail code.
Duan \textit{et al.}~\cite{Duan2010} proposed this $[[8,1]]$ $1$-code for correcting only an AD error, and Ouyang~\cite{Ouyang2021} independently realized this CE code also avoids CC errors.

The codewords of the inner dual-rail code are shown as $\ket{0}_{\textrm{inner}}=\ket{0}\ket{1}$ and $\ket{1}_{\textrm{inner}}=\ket{1}\ket{0}$.
Consequently, we have a family of $[[2(w+1)(w+K),K]]$ CC-AD QEC codes. A general encoding circuit for these codes is shown in Fig.~\ref{fig:enc[2(w+1)(w+K),K]}. Corollary~\ref{Cor: AQEC[[2(w+1)(w+K),K]]} comes from theorem~\ref{Thm:AQEC[[(w+1)(w+K),K]]}.

\begin{figure}[htbp]
    \centering
    \includegraphics[scale=0.2]{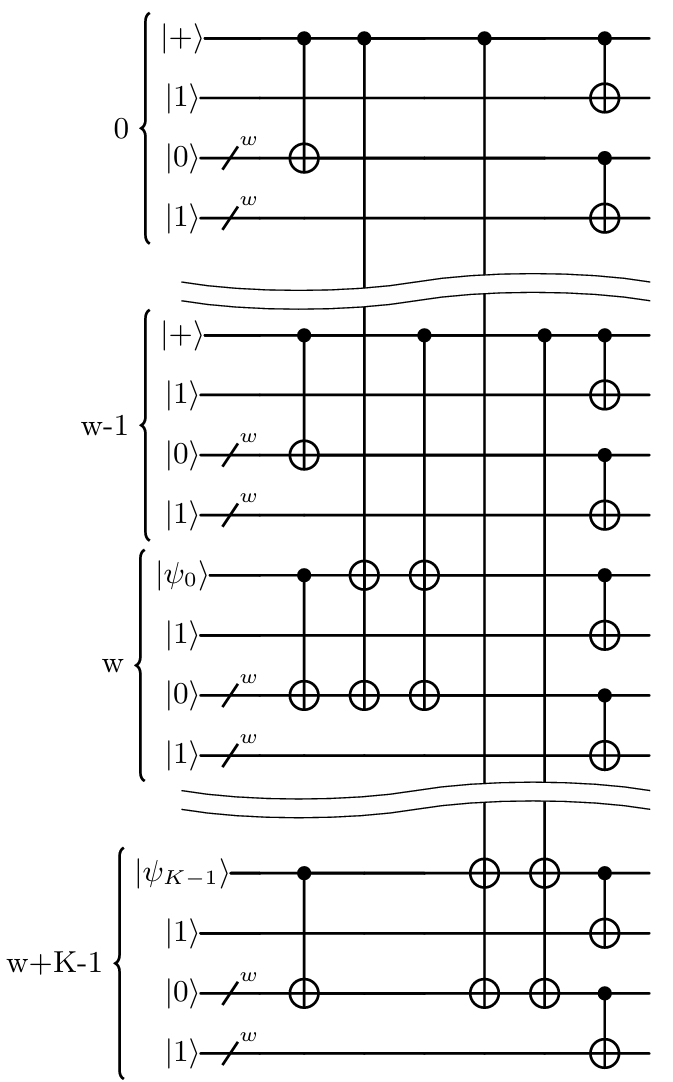}
    \caption{
    Encoding circuit for the $[[2(w+1)(w+K),K]]$ CC-AD QEC codes, which provide a high code rate, immunity to CC errors, and the ability to correct high-weight AD errors. This code encodes $K$ input qubits into $2(w+1)(w+K)$ qubits, protecting against up to weight-$w$ AD errors. The asymptotic code rate is $\frac{1}{2(w+1)}$ as $K$ approaches infinity.
 }\label{fig:enc[2(w+1)(w+K),K]}
\end{figure}

\begin{corollary}\label{Cor: AQEC[[2(w+1)(w+K),K]]} (AQEC conditions for the $[[2(w+1)(w+K),K]]$ CC-AD $w$-code.)
The $[[2(w+1)(w+K),K]]$ CC-AD QEC codewords that concatenate an outer dual-rail code and an inner $[[(w+1)(w+K),K]]$ code are CE states that can correct $w$ AD events.
\end{corollary} 

{\textit{(Alternative Constructions of CC-AD codes.)}}
As mentioned in the Introduction, Duan \textit{et al.}~\cite{Duan2010} showed that concatenating a stabilizer code with a dual-rail code results in an AD code, despite the measurement of stabilizers containing $\hat{X}$ might have potential issues. These AD codes inherently belong to the class of CC codes. This leads to the following theorem.
\begin{theorem}\label{thm:NKD_CC_AD}
   Given an $[[N, K, d]]$ stabilizer code, there exists a $[[2N, K]]$ CC-AD $(d-1)$-code. 
\end{theorem}

The codes constructed from Theorem~\ref{thm:NKD_CC_AD} offer good code rates due to the concatenation of optimal $[[N,K,d]]$ codes with a dual-rail code. However, the encoding circuit and logical operations of an optimal $[[N,K,d]]$ code can be complex, requiring careful selection of codes for practical applications.  Appropriate codes should be chosen for practical use.
In contrast, the encoding circuit of the $[[2(w+1)(w+K),K]]$ CC-AD $w$-code is straightforward, as shown in Fig.~\ref{fig:enc[2(w+1)(w+K),K]}. Both approaches present promising candidates for CC-AD codes.

For the CC-AD codes presented in Corollary~\ref{Cor: AQEC[[2(w+1)(w+K),K]]} and Theorem~\ref{thm:NKD_CC_AD}, AD error correction can be performed using the erasure detection scheme outlined in~\cite{Duan2010}, relying solely on stabilizer measurements.

{\textit{(Conclusion.)}}
We constructed two families of quantum error correction codes: (1) AD QEC codes and (2) CC-AD QEC codes. Our AD QEC code family, designed to correct high-weight AD errors, provides explicit encoding circuits and logical operations. 
By leveraging the blockwise orthogonality between $\ket{0}^{\otimes (w+1)}$ and $\ket{1}^{\otimes (w+1)}$, we provide a clear proof of why these AD QEC codes satisfy approximate Knill-Laflamme QEC conditions. 

The family of CC-AD QEC codes is capable of mitigating both CC errors and excitation loss. Evaluating their performance in long-distance quantum communication, especially under mixed noise conditions, remains an important open question. 

Current decoding techniques might be computationally intensive, and further research is needed to explore optimized or approximate decoding methods suitable for real-time quantum error correction.

\begin{acknowledgments}
CYL was supported by the National Science and Technology Council (NSTC) in Taiwan under Grants Nos. 113-2221-E-A49-114-MY3 and 113-2119-M-A49-008.
\end{acknowledgments}


\newpage
\appendix
\onecolumngrid

\section{Approximate quantum error correction conditions for the $[[(w+1)^2,1]]$ $w$-code.}\label{APP:AQEC[[(w+1)^2,1]]}

Before we prove the approximate quantum error correction conditions for the high-rate $[[(w+1)(w+K), K]]$ Shor codes, we illustrate the main idea of the proof for special cases. 
For the special case $K=1$, our code reduces to the $[[(w+1)^{2}, 1]]$ $w$-code, which is a symmetric Shor code. Suppose that an inner bit-flip  $(w+1)$-qubit repetition code is spanned by two basis vectors $\{\ket{0}^{\otimes (w+1)}, \ket{1}^{\otimes (w+1)}\}$. Denote the $X$-basis states of the inner repetition code by $\ket{\pm}_{\textrm{rep}}^{(w,1)}= \frac{1}{\sqrt{2}}\big(\ket{0}^{\otimes (w+1)}\pm\ket{1}^{\otimes (w+1)}\big)$. Then, the $[[(w+1)^{2}, 1]]$ code is defined by the following basis vectors $\ket{\pm}_{\textrm{AD}}^{(w,1)}\coloneqq(\ket{\pm}_{\textrm{rep}}^{(w,1)})^{\otimes (w+1)}$. On the other hand, the codewords in the $Z$-basis are
\begin{align}\label{eq:codeword_w_1}
    \ket{i}_{\textrm{AD}}^{(w,1)}=&\frac{1}{\sqrt{2}}\big(\ket{+}_{\textrm{AD}}^{(w,1)}+(-1)^{i}\ket{- }_{\textrm{AD}}^{(w,1)}\big),\notag\\
    =&\frac{1}{\sqrt{2^{w}}}\sum_{a\in\{0,1\}^{w}:\atop \wt(a)=0 \mod 2}\bigotimes_{j=0}^{w-1} \ket{a_j}^{\otimes (w+1)}\otimes\ket{i}^{\otimes(w+1)}+\frac{1}{\sqrt{2^{w}}}\sum_{b\in\{0,1\}^{w}:\atop \wt(b)=1 \mod 2}\bigotimes_{j=0}^{w-1} \ket{b_j}^{\otimes (w+1)}\otimes\ket{i'}^{\otimes(w+1)},
\end{align}
where $i' = i+1 \mod 2$.

To illustrate the structure of the codewords, we enumerate representative examples for different cases in Table~\ref{tab:enumerate_codewords_K_1}.
\begin{table}[ht]
    \centering
    \begin{tabular}{||c||c|c||}
        \hline
        $\ket{i}_{\textrm{AD}}^{(w,K)}$ \textbackslash $(w,K)$ & $(1,1)$ & $(2,1)$\\
        \hline
        \hline
        $\ket{0}_{\textrm{AD}}^{(w,K)}$ & $\frac{1}{\sqrt{2}}\left(\ket{0}^{\otimes 4}+\ket{1}^{\otimes 4}\right)$ & $\frac{1}{2}\left(\ket{0}^{\otimes 9}+\ket{1}^{\otimes 6}\ket{0}^{\otimes 3}+\ket{0}^{\otimes 3}\ket{1}^{\otimes 6}+\ket{1}^{\otimes 3}\ket{0}^{\otimes 3}\ket{1}^{\otimes 3}\right)$\\
        \hline
        $\ket{1}_{\textrm{AD}}^{(w,K)}$ & $\frac{1}{\sqrt{2}}\left(\ket{0}^{\otimes 2}\ket{1}^{\otimes 2}+\ket{1}^{\otimes 2}\ket{0}^{\otimes 2}\right)$ & $\frac{1}{2}\left(\ket{1}^{\otimes 9}+\ket{0}^{\otimes 6}\ket{1}^{\otimes 3}+\ket{1}^{\otimes 3}\ket{0}^{\otimes 6}+\ket{0}^{\otimes 3}\ket{1}^{\otimes 3}\ket{0}^{\otimes 3}\right)$\\
        \hline
    \end{tabular}
    \caption{Examples of codewords for the $[[4,1]]$ $1$-code and the $[[9,1]]$ $2$-code.}
    \label{tab:enumerate_codewords_K_1}
\end{table}

Using the codewords presented in Eq.~(\ref{eq:codeword_w_1}) and the AD error set, we prove that the AQEC conditions stated in Lemma~\ref{Lemma: AQEC[[(w+1)^2,1]]} are satisfied.

\begin{lemma}\label{Lemma: AQEC[[(w+1)^2,1]]}(AQEC conditions for the $[[(w+1)^2,1]]$ $w$-code.)
Let $\{\ket{i}_{\textrm{AD}}^{(w,1)}| i\in\{0,1\}\}$ be its codewords. Define the correctable error set $\mathcal{A}= \{\hat{\mathcal{A}}_{k}: k\in\{0,1\}^{(w+1)^2},  \wt(k)\leq w  \}$. Then, for any $\hat{\mathcal{A}}_{k}, \hat{\mathcal{A}}_{\ell}\in\mathcal{A}$,  we have
\begin{align}
\bra{i}_{\textrm{AD}}^{(w,1)}\hat{\mathcal{A}}_{k}^{\dagger}\hat{\mathcal{A}}_{\ell}\ket{j}_{\textrm{AD}}^{(w,1)}=\delta_{ij}C_{k\ell}+O(\gamma^{w+1}),\label{eq:EC_condition}
\end{align}
where $\delta_{ij}$ is the Kronecker delta function, and $C_{k\ell}$ is
a complex number independent of $i$ and $j$.
\end{lemma}

\textbf{Proof.}
Our strategy for proving that the AQEC conditions are satisfied proceeds in three steps:
\begin{itemize}
    \item[(1)] Verify the case $i\neq j$.
    \item[(2)] Verify the case $i= j$ and $k\neq\ell$.
    \item[(3)] Verify the case $i= j$ and $k=\ell$.
\end{itemize}
We begin with the first step, which is straightforward. One can verify that 
\[
\bra{i}_{\textrm{AD}}^{(w,1)}\hat{\mathcal{A}}_{k}^{\dagger}\hat{\mathcal{A}}_{\ell}\ket{j}_{\textrm{AD}}^{(w,1)}=0
\]
for $i\neq j$. This follows from the fact that $\hat{\mathcal{A}}_{k}\ket{0}_{\textrm{AD}}^{(w,1)}$ represents the collapse of $\ket{0}_{\textrm{AD}}^{(w,1)}$ into a superposition of basis states of the form $\bigotimes_{j=0}^{w} \widetilde{\ket{a_j}}^{\otimes(w+1)}$, where each $\widetilde{\ket{a_j}}^{\otimes(w+1)}$ is either $\ket{0}^{\otimes (w+1)}$ or a computational basis vector of Hamming weight at least $1$, indicating $\ket{a_j}^{\otimes(w+1)}=\ket{1}^{\otimes(w+1)}$. This allows us to identify the original logical state. Thus, the inner product vanishes, as required.

The second step proceeds similarly. We again find that the inner product vanishes:
\[
\bra{i}_{\textrm{AD}}^{(w,1)}\hat{\mathcal{A}}_{k}^{\dagger}\hat{\mathcal{A}}_{\ell}\ket{i}_{\textrm{AD}}^{(w,1)}=0
\]
for $k\neq \ell$. This orthogonality arises from the fact that distinct Kraus operators $\hat{\mathcal{A}}_{k}$ and $\hat{\mathcal{A}}_{\ell}$, yielding orthogonal error syndromes in the code space. As a result, the inner product between the resulting error states remains zero.

Finally, the third step is somewhat more involved. Let us define $\bra{0}_{\textrm{AD}}^{(w,1)}\hat{\mathcal{A}}_{k}^{\dagger}\hat{\mathcal{A}}_{k}\ket{0}_{\textrm{AD}}^{(w,1)}=C_{kk}$ for some complex constant $C_{kk}$. It remains to show that
\begin{align}
    \bra{1}_{\textrm{AD}}^{(w,1)}\hat{\mathcal{A}}_{k}^{\dagger} \hat{\mathcal{A}}_{k}\ket{1}_{\textrm{AD}}^{(w,1)}=&C_{kk}+O(\gamma^{w+1}), \label{eq:approximate_2}
\end{align}
where the operator $\hat{\mathcal{A}}_{k}^{\dagger}\hat{\mathcal{A}}_{k}$ takes the tensor-product form
\begin{align}
   \hat{\mathcal{A}}_{k}^{\dagger}\hat{\mathcal{A}}_{k}=\bigotimes_{j=0}^{(w+1)^2-1}\Big( \frac{1}{2}\hat{I}+(-1)^{k_{j}}(\frac{1}{2}\hat{I}-\gamma\ket{1}\bra{1})\Big).\label{eq:AkAK}
\end{align}

By combining Eqs.~(\ref{eq:codeword_w_1}), (\ref{eq:AkAK}), we can evaluate Eq.~(\ref{eq:approximate_2}) explicitly. For the logical zero state, we have: 
\begin{align}
\bra{0}_{\textrm{AD}}^{(w,1)}\hat{\mathcal{A}}_{k}^{\dagger}\hat{\mathcal{A}}_{k}\ket{0}_{\textrm{AD}}^{(w,1)}
=&\frac{1}{2^{w}}  \sum_{a\in\{0,1\}^{w+1}:\atop \wt(a)=0\mod 2 }\prod_{j=0}^{w} \alpha(a_j),
\end{align}
where
\begin{align}
\alpha(a_j)= \bra{a_j}^{\otimes(w+1)}&\left(\bigotimes_{\ell=(w+1)j}^{(w+1)j+w} \hat{\mathcal{A}}_{k_{\ell}}^{\dagger} \hat{\mathcal{A}}_{k_{\ell}}\right)\ket{a_j}^{\otimes(w+1)}.
\end{align}

Similarly, for the logical one state:
\begin{align}
\bra{1}_{\textrm{AD}}^{(w,1)}\hat{\mathcal{A}}_{k}^{\dagger}\hat{\mathcal{A}}_{k}\ket{1}_{\textrm{AD}}^{(w,1)}
=&\frac{1}{{2^{w}}}  \sum_{b\in\{0,1\}^{w+1}:\atop \wt(b)=1\mod 2 }\prod_{j=0}^{w} \alpha(b_j).
\end{align}

Taking the difference, we obtain:
\begin{align*}&2^{w}\left(\bra{0}_{\textrm{AD}}^{(w,1)}\hat{\mathcal{A}}_{k}^{\dagger}\hat{\mathcal{A}}_{k}\ket{0}_{\textrm{AD}}^{(w,1)}- \bra{1}_{\textrm{AD}}^{(w,1)}\hat{\mathcal{A}}_{k}^{\dagger}\hat{\mathcal{A}}_{k}\ket{1}_{\textrm{AD}}^{(w,1)}\right) \notag\\
=& \sum_{a\in\{0,1\}^{w+1}:\atop \wt(a)=0\mod 2 }\prod_{j=0}^{w} \alpha(a_j)
-\sum_{b\in\{0,1\}^{w+1}:\atop \wt(b)=1\mod 2 }\prod_{j=0}^{w} \alpha(b_j),\\
=& \sum_{a\in\{0,1\}^{w}:\atop \wt(a)=0\mod 2 }\prod_{j=0}^{w-1} \alpha(a_j)\alpha(a_{w}=0)
+\sum_{b\in\{0,1\}^{w}:\atop \wt(b)=1\mod 2 }\prod_{j=0}^{w-1} \alpha(b_j)\alpha(a_{w}=1)\\
&-\sum_{a\in\{0,1\}^{w}:\atop \wt(a)=0\mod 2 }\prod_{j=0}^{w-1} \alpha(a_j)\alpha(b_{w}=1)
-\sum_{b\in\{0,1\}^{w}:\atop \wt(b)=1\mod 2 }\prod_{j=0}^{w-1} \alpha(b_j)\alpha(b_{w}=0),\\
=& \sum_{a\in\{0,1\}^{w}:\atop \wt(a)=0\mod 2 }\prod_{j=0}^{w-1} \alpha(a_j)\left(\alpha(a_{w}=0)-\alpha(b_{w}=1)\right)
+\sum_{b\in\{0,1\}^{w}:\atop \wt(b)=1\mod 2 }\prod_{j=0}^{w-1} \alpha(b_j)\left(\alpha(a_{w}=1)-\alpha(b_{w}=0)\right).
\end{align*}

It can be shown that each difference term, such as $\left(\alpha(a_{w}=0)-\alpha(b_{w}=1)\right)$ , is of order $O(\gamma)$. Consequently,
\begin{align*}&2^{w}\left(\bra{0}_{\textrm{AD}}^{(w,1)}\hat{\mathcal{A}}_{k}^{\dagger}\hat{\mathcal{A}}_{k}\ket{0}_{\textrm{AD}}^{(w,1)}- \bra{1}_{\textrm{AD}}^{(w,1)}\hat{\mathcal{A}}_{k}^{\dagger}\hat{\mathcal{A}}_{k}\ket{1}_{\textrm{AD}}^{(w,1)}\right) \notag\\
=& O(\gamma) \sum_{a\in\{0,1\}^{w}:\atop \wt(a)=0\mod 2 }\prod_{j=0}^{w-1} \alpha(a_j)
- O(\gamma) \sum_{b\in\{0,1\}^{w}:\atop \wt(b)=1\mod 2 }\prod_{j=0}^{w-1} \alpha(b_j).
\end{align*}

By iterating this argument, one finds that the total difference is $O(\gamma^{w+1})$, i.e.,
\begin{align*}
    2^{w}\left(\bra{0}_{\textrm{AD}}^{(w,1)}\hat{\mathcal{A}}_{k}^{\dagger}\hat{\mathcal{A}}_{k}\ket{0}_{\textrm{AD}}^{(w,1)}- \bra{1}_{\textrm{AD}}^{(w,1)}\hat{\mathcal{A}}_{k}^{\dagger}\hat{\mathcal{A}}_{k}\ket{1}_{\textrm{AD}}^{(w,1)}\right)=O(\gamma^{w+1}).
\end{align*}
Consequently, the $[[(w+1)^2, 1]]$ $w$-code can approximately correct AD errors of  weight up to $w$.
\hfill$\blacksquare$

\section{Approximate quantum error correction conditions for the $[[(w+1)(w+K),K]]$ $w$-code.}\label{APP:AQEC[[(w+1)(w+K),K]]}
In this section, we prove that the AQEC conditions are satisfied for the $[[(w+1)(w+K),K]]$ $w$-code by adapting the proof used for the $[[(w+1)^2,1]]$ $w$-code presented in Appendix~\ref{APP:AQEC[[(w+1)^2,1]]}. Suppose that an inner bit-flip $(w+1)$-qubit repetition code is spanned by two basis vectors $\{\ket{0}^{\otimes (w+1)}, \ket{1}^{\otimes (w+1)}\}$. Additionally, the tensor product of $K$ individual $(w+1)$-qubit repetition codes is spanned by the following $2^{K}$ basis vectors $\ket{i}_{\textrm{rep}}^{(w,K)}\coloneqq\bigotimes_{j=0}^{K-1}\ket{i_{j}}^{\otimes w+1}$, where the binary string $i=i_{0}\cdots i_{K-1}\in\{0,1\}^{K}$. Denote $\ket{\pm}_{\textrm{rep}}^{(w,1)}= \frac{1}{\sqrt{2}}\big(\ket{0}^{\otimes (w+1)}\pm\ket{1}^{\otimes (w+1)}\big)$ and $\overline{\ket{\pm i}}_{\textrm{rep}}^{(w,K)}\coloneqq\frac{1}{\sqrt{2}}\big(\ket{i}_{\textrm{rep}}^{(w,K)}\pm\ket{i'}_{\textrm{rep}}^{(w,K)}\big)$, where the binary string $i'=i'_{0}\cdots i'_{K-1}\in\{0,1\}^{K}$ and $i_{\ell}'= i_{\ell}+1 \mod 2, \forall \ell \in \{0,\dots,K-1\}$. Then, our $[[(w+1)(w+K),K]]$ code is defined by the following basis vectors $\overline{\ket{\pm i}}_{\textrm{AD}}^{(w,K)}\coloneqq(\ket{\pm}_{\textrm{rep}}^{(w,1)})^{\otimes w}\otimes\overline{\ket{\pm i}}_{\textrm{rep}}^{(w,K)}$. On the other hand, the codewords in the $Z$-basis are
\begin{align}\label{eq:codeword_w_K}
    \ket{i}_{\textrm{AD}}^{(w,K)}=&\frac{1}{\sqrt{2}}\big(\overline{\ket{+i}}_{\textrm{AD}}^{(w,K)}+(-1)^{i}\overline{\ket{-i}}_{\textrm{AD}}^{(w,K)}\big),\notag\\
    =&\frac{1}{\sqrt{2^{w}}}\sum_{a\in\{0,1\}^{w}:\atop \wt(a)=0 \mod 2}\bigotimes_{j=0}^{w-1} \ket{a_j}^{\otimes (w+1)}\otimes\ket{i}_{\textrm{rep}}^{(w,K)}
    +\frac{1}{\sqrt{2^{w}}}\sum_{b\in\{0,1\}^{w}:\atop \wt(b)=1 \mod 2}\bigotimes_{j=0}^{w-1} \ket{b_j}^{\otimes (w+1)}\otimes\ket{i'}_{\textrm{rep}}^{(w,K)}.
\end{align}

To illustrate the structure of the codewords, we enumerate representative examples for different cases in Table~\ref{tab:enumerate_codewords_K}.
\begin{table}[ht]
    \centering
    \begin{tabular}{||c||c||}
        \hline
        $w$ & $\ket{i=i_{0}i_{1}\dots i_{K-1}}_{\textrm{AD}}^{(w,K)}$\\
        \hline
        \hline
        $1$ & $\frac{1}{\sqrt{2}}\left(\ket{0}^{\otimes 2}\ket{i_{0}}^{\otimes 2}\ket{i_{1}}^{\otimes 2}\cdots\ket{i_{K-1}}^{\otimes 2}+\ket{1}^{\otimes 2}\ket{i_{0}'}^{\otimes 2}\ket{i_{1}'}^{\otimes 2}\cdots\ket{i_{K-1}'}^{\otimes 2}\right)$\\
        \hline
        $2$ & $\frac{1}{2}\left(\ket{0}^{\otimes 6}+\ket{1}^{\otimes 6}\right)\otimes\ket{i_{0}}^{\otimes 3}\ket{i_{1}}^{\otimes 3}\cdots\ket{i_{K-1}}^{\otimes 3}+\frac{1}{2}\left(\ket{0}^{\otimes 3}\ket{1}^{\otimes 3}+\ket{1}^{\otimes 3}\ket{0}^{\otimes 3}\right)\otimes\ket{i_{0}'}^{\otimes 3}\ket{i_{1}'}^{\otimes 3}\cdots\ket{i_{K-1}'}^{\otimes 3}$\\
        \hline
    \end{tabular}
    \caption{Examples of codewords for the $[[2(1+K),1]]$ $1$-code and the $[[3(2+K),1]]$ $2$-code.}
    \label{tab:enumerate_codewords_K}
\end{table}

Using the codewords presented in Eq.~(\ref{eq:codeword_w_K}) and the AD error set, we prove that the AQEC conditions stated in Theorem~\ref{Thm:AQEC[[(w+1)(w+K),K]]} are satisfied.

\textbf{Theorem}~\ref{Thm:AQEC[[(w+1)(w+K),K]]} \textrm{(AQEC conditions for the $[[(w+1)(w+K),K]]$ $w$-code.)}
Let $\big\{\ket{i}_{\textrm{AD}}^{(w,K)}| i \in\{0,1\}^{K}\big\}$ be its codewords. Define the correctable error set $\mathcal{A} = \big\{\hat{\mathcal{A}}_{a} : a \in \{0,1\}^{(w+1)(w+K)}, \wt(a) \leq w \big\}$, where $\wt(k)$ denotes the Hamming weight of $k$.
Then, for any $\hat{\mathcal{A}}_{k}, \hat{\mathcal{A}}_{\ell}\in\mathcal{A}$,  we have
\begin{align*}
{\bra{i}}_{\textrm{AD}}^{(w,K)}\hat{\mathcal{A}}_{k}^{\dagger}\hat{\mathcal{A}}_{\ell}{\ket{j}}_{\textrm{AD}}^{(w,K)}
=&\delta_{ij}C_{k\ell}+O(\gamma^{w+1}),
\end{align*}
where $\delta_{ij}$ is the Kronecker delta function, and $C_{k\ell}$ is a complex number independent of these two length-$K$ binary strings $i$ and $j$.

\textbf{Proof.}

As before, the proof proceeds in three steps:
\begin{itemize}
    \item[(1)] Verify the case $i\neq j$.
    \item[(2)] Verify the case $i= j$ and $k\neq\ell$.
    \item[(3)] Verify the case $i= j$ and $k=\ell$.
\end{itemize}
The first and second steps follow straightforwardly, as in the single-logical-qubit case. For the third step, we begin with the similar setup:

Let ${\bra{0}}_{\textrm{AD}}^{(w,K)}\hat{\mathcal{A}}_{k}^{\dagger}\hat{\mathcal{A}}_{k}{\ket{0}}_{\textrm{AD}}^{(w,K)}=C_{kk}$ be for a complex number $C_{kk}$. It remains to show that
\begin{align}
    {\bra{i}}_{\textrm{AD}}^{(w,K)}\hat{\mathcal{A}}_{k}^{\dagger} \hat{\mathcal{A}}_{k}{\ket{i}}_{\textrm{AD}}^{(w,K)}=&C_{kk}+O(\gamma^{w+1}),
\end{align}
where $i \neq 0$ and the operator $\hat{\mathcal{A}}_{k}^{\dagger}\hat{\mathcal{A}}_{k}$ takes the tensor-product form
\begin{align}
   \hat{\mathcal{A}}_{k}^{\dagger}\hat{\mathcal{A}}_{k}=\bigotimes_{j=0}^{(w+1)(w+K)-1}\Big(\frac{1}{2}\hat{I}+(-1)^{k_{j}}(\frac{1}{2}\hat{I}-\gamma\ket{1}\bra{1})\Big).
\end{align}




For the logical zero state, we have:
\begin{align}
    C_{kk}
    =&\frac{1}{ 2^{w}}  \sum_{a_{0},\dots,a_{w-1}\in\{0,1\};\; a_{w}\in\{0,2^{K}-1\}:\atop \wt(a)=0\mod 2 }\prod_{j=0}^{w} \alpha(a_j),
\end{align}
where
\begin{align}
    \alpha(a_j)
    =& \bra{a_j}^{\otimes(w+1)}\left(\bigotimes_{\ell=(w+1)j}^{(w+1)j+w} \hat{\mathcal{A}}_{k_{\ell}}^{\dagger} \hat{\mathcal{A}}_{k_{\ell}}\right)\ket{a_j}^{\otimes(w+1)},\forall j \in\{0,\dots,w-1\},\\
    \alpha(a_{w})
    =& \bra{a_{w}}^{\otimes(w+1)}\left(\bigotimes_{\ell=(w+1)w}^{(w+1)(w+K)-1} \hat{\mathcal{A}}_{k_{\ell}}^{\dagger} \hat{\mathcal{A}}_{k_{\ell}}\right)\ket{a_{w}}^{\otimes(w+1)}.
\end{align}
Similarly, or the logical one state:
\begin{align}
    {\bra{i}}_{\textrm{AD}}^{(w,K)}\hat{\mathcal{A}}_{k}^{\dagger}\hat{\mathcal{A}}_{k}{\ket{i}}_{\textrm{AD}}^{(w,K)}
    =&\frac{1}{{2^{w}}}  \sum_{b_{0},\dots,b_{w-1}\in\{0,1\};\;b_{w}\in\{i,i'\}:\atop \wt(b)=0\mod 2 }\prod_{j=0}^{w} \alpha(b_j),
\end{align}
where $i'=i+1 \mod 2^{K}$, $\forall j \in\{0,\dots,w-1\}$ and 
\begin{align}
    \alpha(b_j)
    =& \bra{b_j}^{\otimes(w+1)}\left(\bigotimes_{\ell=(w+1)j}^{(w+1)j+w} \hat{\mathcal{A}}_{k_{\ell}}^{\dagger} \hat{\mathcal{A}}_{k_{\ell}}\right)\ket{b_j}^{\otimes(w+1)},\\
    \alpha(b_{w})
    =&\bra{b_{w}}^{\otimes(w+1)}\left(\bigotimes_{\ell=(w+1)w}^{(w+1)(w+K)-1} \hat{\mathcal{A}}_{k_{\ell}}^{\dagger} \hat{\mathcal{A}}_{k_{\ell}}\right)\ket{b_{w}}^{\otimes(w+1)}.
\end{align}

Taking the difference, we obtain:
\begin{align*}&2^{w}\left(\overline{\bra{0}}\hat{\mathcal{A}}_{k}^{\dagger}\hat{\mathcal{A}}_{k}\overline{\ket{0}}- \overline{\bra{i}}\hat{\mathcal{A}}_{k}^{\dagger}\hat{\mathcal{A}}_{k}\overline{\ket{i}}\right) \notag\\
=& \sum_{a_{0},\dots,a_{w-1}\in\{0,1\};\; a_{w}\in\{0,2^{K}-1\}:\atop \wt(a)=0\mod 2}\prod_{j=0}^{w} \alpha(a_j)-\sum_{a_{0},\dots,a_{w-1}\in\{0,1\};\;a_{w}\in\{i,i'\}:\atop \wt(a)=0\mod 2 }\prod_{j=0}^{w} \alpha(a_j),\\
=& \sum_{a_{0},\dots,a_{w-1}\in\{0,1\}:\atop \wt(a)=0\mod 2 }\prod_{j=0}^{w-1} \alpha(a_j)\alpha(a_{w}=0)+\sum_{b_{0},\dots,b_{w-1}\in\{0,1\}:\atop \wt(b)=1\mod 2 }\prod_{j=0}^{w-1} \alpha(b_j)\alpha(a_{w}=2^{K}-1)\\
&-\sum_{a_{0},\dots,a_{w-1}\in\{0,1\}:\atop \wt(a)=0\mod 2 }\prod_{j=0}^{w-1} \alpha(a_j)\alpha(b_{w}=i)-\sum_{b_{0},\dots,b_{w-1}\in\{0,1\}:\atop \wt(b)=1\mod 2 }\prod_{j=0}^{w-1} \alpha(b_j)\alpha(b_{w}=i'),\\
=& \sum_{a_{0},\dots,a_{w-1}\in\{0,1\}:\atop \wt(a)=0\mod 2 }\prod_{j=0}^{w-1} \alpha(a_j)\left(\alpha(a_{w}=0)-\alpha(b_{w}=i)\right)+\sum_{b_{0},\dots,b_{w-1}\in\{0,1\}:\atop \wt(b)=1\mod 2 }\prod_{j=0}^{w-1} \alpha(b_j)\left(\alpha(a_{w}=2^{K}-1)-\alpha(b_{w}=i')\right).
\end{align*}

It can be shown that each difference term, such as $\left(\alpha(a_{w}=0)-\alpha(b_{w}=i)\right)$ , is of order $O(\gamma)$. Consequently,
\begin{align*}
2^{w}\left(\overline{\bra{0}}\hat{\mathcal{A}}_{k}^{\dagger}\hat{\mathcal{A}}_{k}\overline{\ket{0}}- \overline{\bra{i}}\hat{\mathcal{A}}_{k}^{\dagger}\hat{\mathcal{A}}_{k}\overline{\ket{i}}\right)
=& O(\gamma) \sum_{a\in\{0,1\}^{w}:\atop \wt(a)=0\mod 2 }\prod_{j=0}^{w-1} \alpha(a_j)
- O(\gamma) \sum_{b\in\{0,1\}^{w}:\atop \wt(b)=1\mod 2 }\prod_{j=0}^{w-1} \alpha(b_j).
\end{align*}

By iterating this argument, one finds that the total difference is $O(\gamma^{w+1})$, i.e.,
\begin{align*}
    2^{w}\left(\overline{\bra{0}}\hat{\mathcal{A}}_{k}^{\dagger}\hat{\mathcal{A}}_{k}\overline{\ket{0}}- \overline{\bra{i}}\hat{\mathcal{A}}_{k}^{\dagger}\hat{\mathcal{A}}_{k}\overline{\ket{i}}\right)=O(\gamma^{w+1}),
\end{align*}
for $i\neq 0$. Consequently, the $[[(w+1)(w+K), K]]$ $w$-code can approximately correct AD errors of weight up to $w$.
\hfill$\blacksquare$

\section{The $[[4,1]]$ $1$-code for AD QEC and the corresponding $[[8,1]]$ $1$-code CC-AD QEC.}\label{APP:[[4,1]]}
This section estimates the overhead of converting a standard AD code into a CE code in the absence of CC errors. We focus on the specific example presented by Ouyang~\cite{Ouyang2021}.

The original $[[4,1]]$ $1$-code, which corresponds to a variant of the Shor code, has logical states defined as follows:
\begin{align*}
    \ket{0}_{\textrm{AD}}^{(1,1)}=&\frac{1}{\sqrt{2}}(\ket{0}^{\otimes 4}+\ket{1}^{\otimes 4}),\\
    \ket{1}_{\textrm{AD}}^{(1,1)}=&\frac{1}{\sqrt{2}}(\ket{0}^{\otimes 2}\ket{1}^{\otimes 2}+\ket{1}^{\otimes 2}\ket{0}^{\otimes 2}).
\end{align*}

Although we take the AD rate $\gamma$ as the physical error rate, the worst-case fidelity between the original and the recovered qubit in the $[[4,1]]$ $1$-code is defined by considering that all AD errors of weight $\geq 2$ are uncorrectable.
\begin{align}
    \mathcal{F}_{\textrm{AD}}^{(1,1)}=\underbrace{(1-\gamma)^{2}}_{\text{weight-0 overlap}}+4\underbrace{[\frac{\gamma(1-\gamma)^{3}}{2}]}_{\text{weight-1 overlap}}.
\end{align}
To assess long-term performance, we consider the effect of $T$ consecutive applications of independent and identical AD channels, each with transmissivity $1-\gamma$. The overall transmissivity becomes $(1-\gamma)^T$. For a fixed damping rate $\gamma$, the QEC scheme remains effective if the worst-case fidelity after error correction satisfies:
\begin{align}
    (1-\gamma)^{2T}+4[\frac{(1-(1-\gamma)^{T})(1-\gamma)^{3T}}{2}]\ge(1-\gamma)^{T}.
\end{align}
Solving this inequality yields a threshold number of time steps,
\begin{align}
    T_{\textrm{th}}\equiv -\frac{\ln{2}}{2\ln{(1-\gamma)}}
\end{align}
beyond which the QEC scheme ceases to provide a net benefit. This bound ensures that the corrected fidelity remains higher than the uncorrected one under repeated channel application.

Additionally, our analysis refines a key estimate in~\cite{Ouyang2021}. Specifically, we find that the infidelity of the $[[8,1]]$ $1$-code is $6\gamma^{2}$, rather than the previously reported $28\gamma^{2}$. For comparison, the infidelity of the original $[[4,1]]$ $1$-code is $5\gamma^{2}$. Hence, the cost of achieving CC-error immunity in the absence of CC errors is an increase in infidelity from $5\gamma^2$ to $6\gamma^2$, due to the use of additional qubits.

\section{Recovery operation of $[[(w+1)(w+K),K]]$ $w$-codes.}\label{APP:decode[[(w+1)(w+K),K]]}
To show that the $[[(w+1)(w+K),K]]$ $w$-code can recover AD errors of weight up to $w$, we observe that the corresponding error states are mutually orthogonal. The recovery operation is constructed accordingly and derived below.

For simplicity, we define the normalized error state and a corresponding projector as
\begin{align}
    \ket{i'}_{(k)}
    =&\frac{\hat{\mathcal{A}}_{k}\ket{i}_{\textrm{AD}}^{(w,K)}}{\sqrt{\bra{i}_{\textrm{AD}}^{(w,K)}\hat{O}_{k}\ket{i}_{\textrm{AD}}^{(w,K)}}},\notag\\
    \hat{P}_{\textrm{AD}}^{(w,K)}
    =&\sum_{\mathclap{\substack{
    i\in\{0,1\}^{K}\\
    k\in\{0,1\}^{(w+1)(w+K)}\\
    1\le \wt(k)\le w \\}}}  \ket{i'}_{(k)}\bra{i'}_{(k)},
\end{align}
where $i=i_{0}\cdots i_{K-1}\in\{0,1\}^{K}$ is the length-$K$ binary string and $k=k_{0}\cdots k_{(w+1)(w+K)-1}\in\{0,1\}^{\otimes (w+1)(w+K)}$ is a binary string representing the error configuration.

Using these definitions, the Kraus operator that recovers the logical state is given by
\begin{align}
    \hat{R}_{\textrm{AD}}^{(w,K)}
    =&\frac{\hat{I}-\hat{P}_{\textrm{AD}}^{(w,K)}}{2}
    +\sum_{\mathclap{\substack{
    i\in\{0,1\}^{K}\\
    k\in\{0,1\}^{(w+1)(w+K)}\\
    1\le \wt(k)\le w \\}}}\ket{i}_{\textrm{AD}}^{(w,K)}\bra{i'}_{(k)}.
\end{align}

Let the encoded logical state be $\ket{\psi}_{\textrm{AD}}^{(w,K)}=\sum_{i\in\{0,1\}^{K}}\alpha_{i}\ket{i}_{\textrm{AD}}^{(w,K)}$, with the corresponding density matrix $\rho=\ket{\psi}_{\textrm{AD}}^{(w,K)}\bra{\psi}_{\textrm{AD}}^{(w,K)}$. After the occurrence of AD errors, the resulting error state has the density matrix
\begin{align}
    \sum_{k\in\{0,1\}^{(w+1)(w+K)}}\hat{\mathcal{A}}_{k}\rho\hat{\mathcal{A}}_{k}^{\dagger}.
\end{align}
Applying the recovery operation yields the final recovered density matrix: 
\begin{align}
    &\sum_{k\in\{0,1\}^{(w+1)(w+K)}}\hat{R}_{\textrm{AD}}^{(w,K)}\hat{\mathcal{A}}_{k}\rho\hat{\mathcal{A}}_{k}^{\dagger}{(\hat{R}_{\textrm{AD}}^{(w,K)})}^{\dagger}.
\end{align}

An explicit example of this recovery procedure is provided in Appendix~\ref{APP:6_2_1_code}.

\section{Analysis of the $[[6,2]]$ $1$-code.}\label{APP:6_2_1_code}
We begin by considering the four codewords of the $[[6,2]]$ $1$-code:
\begin{align}
    \ket{00}_{\textrm{AD}}^{(1,2)}=&\frac{1}{\sqrt{2}}\ket{0}^{\otimes 2}\ket{0}^{\otimes 2}\ket{0}^{\otimes 2}
    +\frac{1}{\sqrt{2}}\ket{1}^{\otimes 2}\ket{1}^{\otimes 2}\ket{1}^{\otimes 2},\notag\\
    \ket{01}_{\textrm{AD}}^{(1,2)}=&\frac{1}{\sqrt{2}}\ket{0}^{\otimes 2}\ket{0}^{\otimes 2}\ket{1}^{\otimes 2}
    +\frac{1}{\sqrt{2}}\ket{1}^{\otimes 2}\ket{1}^{\otimes 2}\ket{0}^{\otimes 2},\notag\\
    \ket{10}_{\textrm{AD}}^{(1,2)}=&\frac{1}{\sqrt{2}}\ket{0}^{\otimes 2}\ket{1}^{\otimes 2}\ket{0}^{\otimes 2}
    +\frac{1}{\sqrt{2}}\ket{1}^{\otimes 2}\ket{0}^{\otimes 2}\ket{1}^{\otimes 2},\notag\\
    \ket{11}_{\textrm{AD}}^{(1,2)}=&\frac{1}{\sqrt{2}}\ket{0}^{\otimes 2}\ket{1}^{\otimes 2}\ket{1}^{\otimes 2}
    +\frac{1}{\sqrt{2}}\ket{1}^{\otimes 2}\ket{0}^{\otimes 2}\ket{0}^{\otimes 2}.
\end{align}

Since this is a weight-$1$ code, we restrict our attention to errors affecting at most one qubit. Consequently, the total number of error events is $\binom{6}{0}+\binom{6}{1}=1+6$ possible events.

For logical indices labeled as $i = i_{0}i_{1}$ and $j = j_{0}j_{1}$, the standard orthogonality condition for quantum error correction requires that for any $i\neq j$,
\[
{\bra{i}}_{\textrm{AD}}^{(1,2)}\hat{\mathcal{A}}_{k}^{\dagger}\hat{\mathcal{A}}_{\ell}{\ket{j}_{\textrm{AD}}^{(1,2)}}=0
\]
since the minimal Hamming distance between distinct codewords is at least two qubits—beyond our single-qubit error set.

Similarly, we must verify that for any fixed codeword (say, for $i$) and for all distinct error events ($k \neq \ell$),
\[
{\bra{i}}_{\textrm{AD}}^{(1,2)}\hat{\mathcal{A}}_{k}^{\dagger}\hat{\mathcal{A}}_{\ell}{\ket{i}}_{\textrm{AD}}^{(1,2)}=0
\]
Without loss of generality, consider $i = 00$. There are $42$ ordered pairs $(k,\ell)$ with $k\neq \ell$ (since $7\times6=42$), but it suffices to check only the cases with $k>\ell$, as the remaining cases are their conjugates. In each such pair, at least one qubit undergoes an $\hat{A}{1}$ error in the $k$-event and an $\hat{A}{0}$ error in the $\ell$-event, ensuring that the resulting states are orthogonal. For instance:
\begin{itemize}
    \item For $k=000001$ and $\ell=000000$, we have
    \begin{align} {\bra{00}}_{\textrm{AD}}^{(1,2)}\hat{\mathcal{A}}_{000001}^{\dagger}\hat{\mathcal{A}}_{000000}{\ket{00}}_{\textrm{AD}}^{(1,2)}=&\left(\frac{\sqrt{1-\gamma}^{5}*\gamma}{\sqrt{2}}\bra{1}^{\otimes 5}\bra{0}\right)\left(\frac{1}{\sqrt{2}}\ket{0}^{\otimes 6}+\frac{(1-\gamma)^{3}}{\sqrt{2}}\ket{1}^{\otimes 6}\right)=0. \end{align}
    \item For $k=000010$ and $\ell=000001$, we obtain 
    \begin{align} {\bra{00}}_{\textrm{AD}}^{(1,2)}\hat{\mathcal{A}}_{000010}^{\dagger}\hat{\mathcal{A}}_{000001}{\ket{00}}_{\textrm{AD}}^{(1,2)}=&\left(\frac{\sqrt{1-\gamma}^{5}*\gamma}{\sqrt{2}}\bra{1}^{\otimes 4}\bra{0}\bra{1}\right)\left(\frac{\sqrt{1-\gamma}^{5}*\gamma}{\sqrt{2}}\ket{1}^{\otimes 5}\ket{0}\right)=0. \end{align}
    Analogous arguments apply to the remaining cases.
\end{itemize}

Next, let us denote ${\bra{0}}_{\textrm{AD}}^{(1,2)}\hat{\mathcal{A}}_{k}^{\dagger}\hat{\mathcal{A}}_{k}{\ket{0}}_{\textrm{AD}}^{(1,2)}=C_{kk}$ with $C_{kk}$ being a complex constant. For the diagonal terms (i.e. $k=\ell$) and any codeword $\ket{i}_{\textrm{AD}}^{(1,2)}$ (with $i\neq 0$), we wish to demonstrate that
\begin{align}
    {\bra{i}}_{\textrm{AD}}^{(1,2)}\hat{\mathcal{A}}_{k}^{\dagger} \hat{\mathcal{A}}_{k}{\ket{i}}_{\textrm{AD}}^{(1,2)}=&C_{kk}+O(\gamma^{2}),
\end{align}
where $i \neq 0$ and 
\begin{align}
   \hat{\mathcal{A}}_{k}^{\dagger}\hat{\mathcal{A}}_{k}=\bigotimes_{j=0}^{5}\Big(\frac{1}{2}\hat{I}+(-1)^{k_{j}}(\frac{1}{2}\hat{I}-\gamma\ket{1}\bra{1})\Big).
\end{align}

We now analyze the states $\hat{\mathcal{A}}_{k}\ket{i}_{\textrm{AD}}^{(1,2)}$ for each $i\in \{00,01,10,11\}$ and for each of the 7 possible error patterns indexed by $k$. The results are summarized in Table~\ref{table:example_6_2_1}.
\begin{table}[ht]
\begin{adjustbox}{width=0.9\columnwidth,center}
\begin{tabular}{||c | c | c | c | c | c ||} 
\hline
 \textrm{k}\textbackslash \textrm{i} & $00$ & $01$\\ [0.5ex] 
 \hline
 $000000$ & $\left(\frac{1}{\sqrt{2}}\ket{0}^{\otimes 6}+\frac{(1-\gamma)^{3}}{\sqrt{2}}\ket{1}^{\otimes 6}\right)$ & $\left(\frac{1-\gamma}{\sqrt{2}}\ket{0}^{\otimes 4}\ket{1}^{\otimes 2}+\frac{(1-\gamma)^{2}}{\sqrt{2}}\ket{1}^{\otimes 4}\ket{0}^{\otimes 2}\right)$\\  
 \hline
 $000001$ & $\left(\frac{\gamma*\sqrt{1-\gamma}^5}{\sqrt{2}}\ket{1}^{\otimes 5}\ket{0}\right)$ & $\left(\frac{\gamma*\sqrt{1-\gamma}}{\sqrt{2}}\ket{0}^{\otimes 4}\ket{1}\ket{0}\right)$\\  
 \hline
 $000010$ & $\left(\frac{\gamma*\sqrt{1-\gamma}^5}{\sqrt{2}}\ket{1}^{\otimes 4}\ket{0}\ket{1}\right)$ & $\left(\frac{\gamma*\sqrt{1-\gamma}}{\sqrt{2}}\ket{0}^{\otimes 5}\ket{1}\right)$\\  
 \hline
 $000100$ & $\left(\frac{\gamma*\sqrt{1-\gamma}^5}{\sqrt{2}}\ket{1}^{\otimes 3}\ket{0}\ket{1}^{\otimes 2}\right)$ & $\left(\frac{\gamma*\sqrt{1-\gamma}^{3}}{\sqrt{2}}\ket{1}^{\otimes 3}\ket{0}^{\otimes 3}\right)$\\  
 \hline
 $001000$ & $\left(\frac{\gamma*\sqrt{1-\gamma}^5}{\sqrt{2}}\ket{1}^{\otimes 2}\ket{0}\ket{1}^{\otimes 3}\right)$ & $\left(\frac{\gamma*\sqrt{1-\gamma}^{\otimes 3}}{\sqrt{2}}\ket{1}^{\otimes 2}\ket{0}\ket{1}\ket{0}^{\otimes 2}\right)$ \\  
 \hline
 $010000$ & $\left(\frac{\gamma*\sqrt{1-\gamma}^5}{\sqrt{2}}\ket{1}\ket{0}\ket{1}^{\otimes 4}\right)$ & $\left(\frac{\gamma*\sqrt{1-\gamma}^{3}}{\sqrt{2}}\ket{1}\ket{0}\ket{1}^{\otimes 2}\ket{0}^{\otimes 2}\right)$ \\  
 \hline
 $100000$ & $\left(\frac{\gamma*\sqrt{1-\gamma}^5}{\sqrt{2}}\ket{0}\ket{1}^{\otimes 5}\right)$ & $\left(\frac{\gamma*\sqrt{1-\gamma}^{3}}{\sqrt{2}}\ket{0}\ket{1}^{\otimes 3}\ket{0}^{\otimes 2}\right)$\\  
 \hline
 \hline
 \textrm{k}\textbackslash \textrm{i} & $10$ & $11$\\ [0.5ex] 
 \hline
 $000000$ & $\left(\frac{1-\gamma}{\sqrt{2}}\ket{0}^{\otimes 2}\ket{1}^{\otimes 2}\ket{0}^{\otimes 2}+\frac{(1-\gamma)^{2}}{\sqrt{2}}\ket{1}^{\otimes 2}\ket{0}^{\otimes 2}\ket{1}^{\otimes 2}\right)$ & $\left(\frac{(1-\gamma)^{2}}{\sqrt{2}}\ket{0}^{\otimes 2}\ket{1}^{\otimes 4}+\frac{1-\gamma}{\sqrt{2}}\ket{1}^{\otimes 2}\ket{0}^{\otimes 4}\right)$\\  
 \hline
 $000001$ & $\left(\frac{\gamma*\sqrt{1-\gamma}^{3}}{\sqrt{2}}\ket{1}^{\otimes 2}\ket{0}^{\otimes 2}\ket{1}\ket{0}\right)$ & $\left(\frac{\gamma*\sqrt{1-\gamma}^{3}}{\sqrt{2}}\ket{0}^{\otimes 2}\ket{1}^{\otimes 3}\ket{0}\right)$\\  
 \hline
 $000010$ & $\left(\frac{\gamma*\sqrt{1-\gamma}^{3}}{\sqrt{2}}\ket{1}^{\otimes 2}\ket{0}^{\otimes 3}\ket{1}\right)$ & $\left(\frac{\gamma*\sqrt{1-\gamma}^{3}}{\sqrt{2}}\ket{0}^{\otimes 2}\ket{1}^{\otimes 2}\ket{0}\ket{1}\right)$\\  
 \hline
 $000100$ & $\left(\frac{\gamma*\sqrt{1-\gamma}}{\sqrt{2}}\ket{0}^{\otimes 2}\ket{1}\ket{0}^{\otimes 3}\right)$ & $\left(\frac{\gamma*\sqrt{1-\gamma}^{3}}{\sqrt{2}}\ket{0}^{\otimes 2}\ket{1}\ket{0}\ket{1}^{\otimes 2}\right)$\\  
 \hline
 $001000$ & $\left(\frac{\gamma*\sqrt{1-\gamma}}{\sqrt{2}}\ket{0}^{\otimes 3}\ket{1}\ket{0}^{\otimes 2}\right)$ & $\left(\frac{\gamma*\sqrt{1-\gamma}^{3}}{\sqrt{2}}\ket{0}^{\otimes 3}\ket{1}^{\otimes 3}\right)$ \\  
 \hline
 $010000$ & $\left(\frac{\gamma*\sqrt{1-\gamma}^{3}}{\sqrt{2}}\ket{1}\ket{0}^{\otimes 3}\ket{1}^{\otimes 2}\right)$ & $\left(\frac{\gamma*\sqrt{1-\gamma}}{\sqrt{2}}\ket{1}\ket{0}^{\otimes 5}\right)$ \\  
 \hline
 $100000$ & $\left(\frac{\gamma*\sqrt{1-\gamma}^{3}}{\sqrt{2}}\ket{1}^{\otimes 2}\ket{0}^{\otimes 2}\ket{1}^{\otimes 2}\right)$ & $\left(\frac{\gamma*\sqrt{1-\gamma}}{\sqrt{2}}\ket{0}\ket{1}\ket{0}^{\otimes 4}\right)$\\  
 \hline
\end{tabular}
\end{adjustbox}
\caption{The corresponding terms of the AQEC conditions for the $[[6, 2]]$ $1$-codes in this work.
}
\label{table:example_6_2_1}
\end{table}

From the table, we observe that
  \[
    C_{kk} = \left\{\begin{array}{lr}
        \frac{1}{2}(1+(1-\gamma)^{6}), & \text{for } k=000000\\
        \frac{\gamma^2*(1-\gamma)^{5}}{2}, & \text{otherwise}
        \end{array}\right\}.
  \]

Since the contributions from every state with $k\neq 000000$ are of order $O(\gamma^{2})$, it suffices to verify the condition for $k=000000$. In this case, we require \begin{align} \frac{1}{2}\Bigl(1+(1-\gamma)^{6}\Bigr) - \frac{1}{2}\Bigl((1-\gamma)^{2}+(1-\gamma)^{4}\Bigr) = O(\gamma^{2}), \end{align} which is clearly satisfied.

Furthermore, we can extract the error syndrome for this error set. The first step is to measure three stabilizers, \{$Z_0Z_1$, $Z_2Z_3$, $Z_4Z_5$\}. This is implemented by applying CNOT gates with controls on qubits $\{0, 2, 4\}$ and targets on qubits $\{1, 3, 5\}$, followed by measuring qubits $\{1, 3, 5\}$. In this way, the remaining qubits $\{0, 2, 4\}$ collapse into states that correspond to the syndrome outcomes shown in Table~\ref{table:6_2_1_a}.
\begin{table}[ht]
\begin{adjustbox}{width=0.9\columnwidth,center}
\begin{tabular}{||c |c | c | c | c | c | c ||} 
\hline
 \textrm{k}\textbackslash \textrm{i} & \textrm{syndrome} & $00$ & $01$\\ [0.5ex] 
 \hline
 $000000$ & $000$ & $\left(\frac{1}{\sqrt{2}}\ket{0}^{\otimes 3}+\frac{(1-\gamma)^{3}}{\sqrt{2}}\ket{1}^{\otimes 3}\right)$ & $\left(\frac{1-\gamma}{\sqrt{2}}\ket{0}^{\otimes 2}\ket{1}+\frac{(1-\gamma)^{2}}{\sqrt{2}}\ket{1}^{\otimes 2}\ket{0}\right)$\\  
 \hline
 $000001$ & $001$ & $\left(\frac{\gamma*\sqrt{1-\gamma}^5}{\sqrt{2}}\ket{1}^{\otimes 3}\right)$ & $\left(\frac{\gamma*\sqrt{1-\gamma}}{\sqrt{2}}\ket{0}^{\otimes 2}\ket{1}\right)$\\  
 \hline
 $000010$ & $001$ & $\left(\frac{\gamma*\sqrt{1-\gamma}^5}{\sqrt{2}}\ket{1}^{\otimes 2}\ket{0}\right)$ & $\left(\frac{\gamma*\sqrt{1-\gamma}}{\sqrt{2}}\ket{0}^{\otimes 3}\right)$\\  
 \hline
 $000100$ & $010$ & $\left(\frac{\gamma*\sqrt{1-\gamma}^5}{\sqrt{2}}\ket{1}^{\otimes 3}\right)$ & $\left(\frac{\gamma*\sqrt{1-\gamma}^{3}}{\sqrt{2}}\ket{1}^{\otimes 2}\ket{0}\right)$\\  
 \hline
 $001000$ & $010$ & $\left(\frac{\gamma*\sqrt{1-\gamma}^5}{\sqrt{2}}\ket{1}\ket{0}\ket{1}\right)$ & $\left(\frac{\gamma*\sqrt{1-\gamma}^{3}}{\sqrt{2}}\ket{1}\ket{0}^{\otimes 2}\right)$ \\  
 \hline
 $010000$ & $100$ & $\left(\frac{\gamma*\sqrt{1-\gamma}^5}{\sqrt{2}}\ket{1}^{\otimes 3}\right)$ & $\left(\frac{\gamma*\sqrt{1-\gamma}^{3}}{\sqrt{2}}\ket{1}^{\otimes 2}\ket{0}\right)$ \\  
 \hline
 $100000$ & $100$ & $\left(\frac{\gamma*\sqrt{1-\gamma}^5}{\sqrt{2}}\ket{0}\ket{1}^{\otimes 2}\right)$ & $\left(\frac{\gamma*\sqrt{1-\gamma}^{3}}{\sqrt{2}}\ket{0}\ket{1}\ket{0}\right)$\\  
 \hline
 \hline
 \textrm{k}\textbackslash \textrm{i} & \textrm{syndrome} & $10$ & $11$\\ [0.5ex] 
 \hline
 $000000$ & $000$ & $\left(\frac{1-\gamma}{\sqrt{2}}\ket{0}\ket{1}\ket{0}+\frac{(1-\gamma)^{2}}{\sqrt{2}}\ket{1}\ket{0}\ket{1}\right)$ & $\left(\frac{(1-\gamma)^{2}}{\sqrt{2}}\ket{0}\ket{1}^{\otimes 2}+\frac{1-\gamma}{\sqrt{2}}\ket{1}\ket{0}^{\otimes 2}\right)$\\  
 \hline
 $000001$ & $001$ & $\left(\frac{\gamma*\sqrt{1-\gamma}^{3}}{\sqrt{2}}\ket{1}\ket{0}\ket{1}\right)$ & $\left(\frac{\gamma*\sqrt{1-\gamma}^{3}}{\sqrt{2}}\ket{0}\ket{1}^{\otimes 2}\right)$\\  
 \hline
 $000010$ & $001$ & $\left(\frac{\gamma*\sqrt{1-\gamma}^{3}}{\sqrt{2}}\ket{1}\ket{0}^{\otimes 2}\right)$ & $\left(\frac{\gamma*\sqrt{1-\gamma}^{3}}{\sqrt{2}}\ket{0}\ket{1}\ket{0}\right)$\\  
 \hline
 $000100$ & $010$ & $\left(\frac{\gamma*\sqrt{1-\gamma}}{\sqrt{2}}\ket{0}\ket{1}\ket{0}\right)$ & $\left(\frac{\gamma*\sqrt{1-\gamma}^{3}}{\sqrt{2}}\ket{0}\ket{1}^{\otimes 2}\right)$\\  
 \hline
 $001000$ & $010$ & $\left(\frac{\gamma*\sqrt{1-\gamma}}{\sqrt{2}}\ket{0}^{\otimes 3}\right)$ & $\left(\frac{\gamma*\sqrt{1-\gamma}^{3}}{\sqrt{2}}\ket{0}^{\otimes 2}\ket{1}\right)$ \\  
 \hline
 $010000$ & $100$ & $\left(\frac{\gamma*\sqrt{1-\gamma}^{3}}{\sqrt{2}}\ket{1}\ket{0}\ket{1}\right)$ & $\left(\frac{\gamma*\sqrt{1-\gamma}}{\sqrt{2}}\ket{1}\ket{0}^{\otimes 2}\right)$ \\  
 \hline
 $100000$ & $100$ & $\left(\frac{\gamma*\sqrt{1-\gamma}^{3}}{\sqrt{2}}\ket{1}\ket{0}\ket{1}\right)$ & $\left(\frac{\gamma*\sqrt{1-\gamma}}{\sqrt{2}}\ket{0}^{\otimes 3}\right)$\\
 \hline
\end{tabular}
\end{adjustbox}
\caption{The lookup table for the output state corresponding to the syndrome.}
\label{table:6_2_1_a}
\end{table}

There are four recorded syndromes $\{000,001,010,100\}$. Conditional on the syndrome, the recovery operations are applied as follows:
\begin{itemize}
    \item[000] Apply two CNOT gates defined by the operator 
    \[(\ket{0}\bra{0})_{0}\otimes\hat{I}_{2}\hat{I}_{4}+(\ket{1}\bra{1})_{0}\otimes\hat{X}_{2}\hat{X}_{4}.\] This operation recovers the desired state on the last two qubits, up to corrections of order $O(\gamma^{2})$ among the logical codewords $\{00,01,10,11\}$.
    \item[100] Discard qubit 0, then apply the Pauli $\hat{X}$ operators on qubits 2 and 4, $X_2X_4$, to obtain an approximate recovery of the state. Finally, apply an artificial amplitude-damping operation,
    \[\hat{A}'_{0}=\ket{0}\bra{0}+(1-\gamma)\ket{1}\bra{1},\]
    on the remaining two qubits to correct an undesired $Y$-rotation. As discussed in the main text, the physical implementation of the artificial AD operator requires only a controlled-$\hat{Y}$ rotation with an ancillary qubit initialized to $\ket{0}$, followed by a post-selected measurement of this ancillary qubit in the $Z$-basis $\ket{0}$.
    \item[001] Discard qubit 4 and then apply $X_0X_2$. Next, apply the artificial amplitude-damping operator
    \[\hat{A}'_{0}=\ket{0}\bra{0}+\sqrt{1-\gamma}\ket{1}\bra{1}\] 
    to qubits 0 and 2 to correct for an undesired $Y$-rotation. Finally, apply a CNOT gate defined as
    \[\hat{I}_{0}\otimes(\ket{0}\bra{0})_{2}+\hat{X}_{0}\otimes(\ket{1}\bra{1})_{2},\]
    which recovers the desired state.
    \item[010] Discard qubit 2 and then apply $X_0X_4$. Next, apply the artificial amplitude-damping operator
    \[\hat{A}'_{0}=\ket{0}\bra{0}+\sqrt{1-\gamma}\ket{1}\bra{1}\] 
    to qubits 0 and 4 to correct for an undesired $Y$-rotation. Finally, apply a CNOT gate defined as
    \[(\ket{0}\bra{0})_{0}\otimes\hat{I}_{4}+(\ket{1}\bra{1})_{0}\otimes\hat{X}_{4},\]
    to obtain the desired state.
\end{itemize}

After applying the decoding circuit, the output states are given in Table~\ref{table:6_2_1_b}. In this table, each row corresponds to an error pattern (indexed by $k$) and its associated syndrome, while the columns display the decoded states for the logical codewords.
\begin{table}[ht]
\begin{adjustbox}{width=0.6\columnwidth,center}
\begin{tabular}{||c |c | c | c | c | c | c ||} 
\hline
 \textrm{k}\textbackslash \textrm{i} & \textrm{syndrome} & $00$ & $01$\\ [0.5ex] 
 \hline
 $000000$ & $000$ & $\left(\frac{2-3\gamma+O(\gamma^{2})}{\sqrt{2}}\ket{0}^{\otimes 2}\right)$ & $\left(\frac{2-3\gamma+O(\gamma^{2})}{\sqrt{2}}\ket{0}\ket{1}\right)$\\  
 \hline
 $000001$ & $001$ & $\left(\frac{\gamma*\sqrt{1-\gamma}^5}{\sqrt{2}}\ket{0}^{\otimes 2}\right)$ & $\left(\frac{\gamma*\sqrt{1-\gamma}^{5}}{\sqrt{2}}\ket{0}\ket{1}\right)$\\  
 \hline
 $000010$ & $001$ & $\left(\frac{\gamma*\sqrt{1-\gamma}^5}{\sqrt{2}}\ket{0}^{\otimes 2}\right)$ & $\left(\frac{\gamma*\sqrt{1-\gamma}^{5}}{\sqrt{2}}\ket{0}\ket{1}\right)$\\  
 \hline
 $000100$ & $010$ & $\left(\frac{\gamma*\sqrt{1-\gamma}^5}{\sqrt{2}}\ket{0}^{\otimes 2}\right)$ & $\left(\frac{\gamma*\sqrt{1-\gamma}^{5}}{\sqrt{2}}\ket{0}\ket{1}\right)$\\  
 \hline
 $001000$ & $010$ & $\left(\frac{\gamma*\sqrt{1-\gamma}^5}{\sqrt{2}}\ket{0}^{\otimes 2}\right)$ & $\left(\frac{\gamma*\sqrt{1-\gamma}^{5}}{\sqrt{2}}\ket{0}\ket{1}\right)$ \\  
 \hline
 $010000$ & $100$ & $\left(\frac{\gamma*\sqrt{1-\gamma}^5}{\sqrt{2}}\ket{0}^{\otimes 2}\right)$ & $\left(\frac{\gamma*\sqrt{1-\gamma}^{5}}{\sqrt{2}}\ket{0}\ket{1}\right)$ \\  
 \hline
 $100000$ & $100$ & $\left(\frac{\gamma*\sqrt{1-\gamma}^5}{\sqrt{2}}\ket{0}^{\otimes 2}\right)$ & $\left(\frac{\gamma*\sqrt{1-\gamma}^{5}}{\sqrt{2}}\ket{0}\ket{1}\right)$\\  
 \hline
 \hline
 \textrm{k}\textbackslash \textrm{i} & \textrm{syndrome} & $10$ & $11$\\ [0.5ex] 
 \hline
 $000000$ & $000$ & $\left(\frac{2-3\gamma+O(\gamma^{2})}{\sqrt{2}}\ket{1}\ket{0}\right)$ & $\left(\frac{2-3\gamma+O(\gamma^{2})}{\sqrt{2}}\ket{1}^{\otimes 2}\right)$\\  
 \hline
 $000001$ & $001$ & $\left(\frac{\gamma*\sqrt{1-\gamma}^{4}}{\sqrt{2}}\ket{1}\ket{0}\right)$ & $\left(\frac{\gamma*\sqrt{1-\gamma}^{4}}{\sqrt{2}}\ket{1}^{\otimes 2}\right)$\\  
 \hline
 $000010$ & $001$ & $\left(\frac{\gamma*\sqrt{1-\gamma}^{4}}{\sqrt{2}}\ket{1}\ket{0}\right)$ & $\left(\frac{\gamma*\sqrt{1-\gamma}^{4}}{\sqrt{2}}\ket{1}^{\otimes 2}\right)$\\  
 \hline
 $000100$ & $010$ & $\left(\frac{\gamma*\sqrt{1-\gamma}^{5}}{\sqrt{2}}\ket{1}\ket{0}\right)$ & $\left(\frac{\gamma*\sqrt{1-\gamma}^{4}}{\sqrt{2}}\ket{1}^{\otimes 2}\right)$\\  
 \hline
 $001000$ & $010$ & $\left(\frac{\gamma*\sqrt{1-\gamma}^{5}}{\sqrt{2}}\ket{1}\ket{0}\right)$ & $\left(\frac{\gamma*\sqrt{1-\gamma}^{4}}{\sqrt{2}}\ket{1}^{\otimes 2}\right)$ \\  
 \hline
 $010000$ & $100$ & $\left(\frac{\gamma*\sqrt{1-\gamma}^{5}}{\sqrt{2}}\ket{1}\ket{0}\right)$ & $\left(\frac{\gamma*\sqrt{1-\gamma}^{5}}{\sqrt{2}}\ket{1}^{\otimes 2}\right)$ \\  
 \hline
 $100000$ & $100$ & $\left(\frac{\gamma*\sqrt{1-\gamma}^{5}}{\sqrt{2}}\ket{1}\ket{0}\right)$ & $\left(\frac{\gamma*\sqrt{1-\gamma}^{5}}{\sqrt{2}}\ket{1}^{\otimes 2}\right)$\\
 \hline
\end{tabular}
\end{adjustbox}
\caption{The lookup table for the resulting state after recovery operations corresponding to the syndrome.}
\label{table:6_2_1_b}
\end{table}

Since the difference between these coefficients, $\frac{\gamma*\sqrt{1-\gamma}^{4}}{\sqrt{2}}\approx\frac{\gamma+O(\gamma^{2})}{\sqrt{2}}$ and $\frac{\gamma*\sqrt{1-\gamma}^{5}}{\sqrt{2}}\approx\frac{\gamma+O(\gamma^{2})}{\sqrt{2}}$, is negligible for $w=1$, applying the re-encoding circuit after decoding successfully recovers the logical states.

\end{document}